\documentclass[aps,prx,twocolumn,superscriptaddress,longbibliography,floatfix]{revtex4-2}

\usepackage{graphicx}% Include figure files
\usepackage[braket, qm]{qcircuit}
\usepackage{amsthm}
\usepackage{amssymb}
\usepackage{mathtools}
\usepackage[super]{nth}

\graphicspath{{Figures/}}
\usepackage{orcidlink}
\hypersetup{hidelinks}

\begin{document}

\title{Error-Corrected QRAM Queries with One-Fault-Tolerant \texttt{CSWAP} Gates}

% Yale combined affiliation (for authors with Physics and/or Applied Physics)
\author{Shifan Xu\orcidlink{0009-0005-9103-228X}}
\affiliation{Department of Applied Physics, Yale University, New Haven, CT 06520, USA}

\author{Connor T.\ Hann\orcidlink{0000-0003-0665-7161}}
\affiliation{Amazon Center for Quantum Computing, Pasadena, CA 91125, USA}

\author{Nathan Wiebe\orcidlink{0000-0001-6047-0547}}
\affiliation{Department of Computer Science, University of Toronto, Toronto, ON M5S 1J7, Canada}
\affiliation{Pacific Northwest National Laboratory, Richland, WA, USA}

\author{Steven M.\ Girvin\orcidlink{0000-0002-6470-5494}}
\affiliation{Department of Physics, Yale University, New Haven, CT 06520, USA}
\affiliation{Department of Applied Physics, Yale University, New Haven, CT 06520, USA}

\author{Yongshan Ding\orcidlink{0000-0002-2338-1315}}
\affiliation{Department of Computer Science, Yale University, New Haven, CT 06520, USA}
\affiliation{Department of Applied Physics, Yale University, New Haven, CT 06520, USA}
\date{\today}% It is always \today, today,

\begin{abstract}
Quantum random-access memory (QRAM) enables coherent queries to classical data. However, the \(O(N)\) qubits required for a memory with \(N=2^n\) addresses make QRAM appear vulnerable to noise. Though bucket-brigade routing provides intrinsic noise resilience to reduce the error in QRAM, improving the fidelity of QRAM queries is still an important and challenging problem. We introduce a low-overhead protection framework for bucket-brigade QRAM with two complementary components. First, we propose a one-fault-tolerant repeat-until-success logical controlled-SWAP gate using the \([[4,2,2]]\) code under erasure noise channel. Interleaved erasure checks and local recovery correct every single erasure, so the leading logical failure probability is second order in the physical erasure error rates. Second, we exploit validity constraints of bucket-brigade routing to define local parity checks directly on the router states, without adding extra data qubits to create code redundancy. Under bit-flip noise, minimum-weight perfect matching with noiseless syndrome measurements corrects single interior faults, while open spatial boundaries retain a first-order contribution in the residual bit-flip rate. Monte Carlo simulations confirm quadratic suppression for the protected controlled-SWAP and improved depth scaling for decoded QRAM router-state snapshots. The two codes admit a concatenated realization in which the inner code supplies erasure protection and the outer code optionally corrects interior logical bit flips. Together, these results show how routing constraints and a protected non-Clifford routing gate can improve QRAM reliability under biased and erasure-dominant noise.

\begin{description}
\item[Keywords]
quantum random-access memory, quantum error correction, fault tolerance
\end{description}
\end{abstract}

\maketitle

\section{Introduction}
\label{sec:intro}

Many modern quantum algorithms assume coherent oracle access to classical data. Quantum random-access memory (QRAM) provides an explicit architecture for realizing this access: it maps a superposition of addresses to the corresponding data in a single coherent query~\cite{giovannetti2008architectures,giovannetti2008quantum}. Coherent data access is used in quantum simulation, linear-system solving, and machine learning, among other applications~\cite{babbush2018encoding,harrow2009quantum,biamonte2017quantum}. Its usefulness, however, depends on whether a large memory can be queried with sufficiently low latency and error~\cite{jaques2023qram}.

QRAM architecture design offers tradeoffs between hardware size, query latency, and bandwidth. Hybrid architectures combine different memory-access schemes to balance space and time costs~\cite{xu2023systems}, while Fat-Tree QRAM pipelines parallel queries to increase bandwidth~\cite{xu2025fattree}. Circuit-based data lookup also permits tradeoffs between ancillary qubits and non-Clifford gates~\cite{low2018trading}.

Bucket-brigade (BB) QRAM is a leading router-based QRAM construction. A size-\(N\) memory uses a binary tree with \(n=\log_2 N\) router levels, giving \(O(n)\) query depth with \(O(N)\) routers~\cite{giovannetti2008architectures,di2020fault,arunachalam2015robustness}. This linear hardware cost is already substantial. Noise creates a second challenge: faults may occur throughout a large circuit and may erase any algorithmic advantage provided by an ideal oracle~\cite{regev2008impossibility,shenvi2003effects}. BB QRAM nevertheless resists such noise: because a query activates only a small part of the tree in each address branch, its infidelity can scale with tree depth rather than with the total number of routers~\cite{hann}. Mehta, Lee, and Jiang extended the noise-resilience analysis to initialization, spatially correlated, and coherent errors, and developed QRAM randomized compiling~\cite{mehta2024analysis}. We use the routing constraints for active, low-overhead error correction.

A natural solution, generic quantum error correction (QEC), is poorly matched to this setting. Encoding every router and implementing every controlled-SWAP (\texttt{CSWAP}) gate fault tolerantly in a universal architecture would multiply an already \(O(N)\) hardware footprint. Moreover, \texttt{CSWAP} is the only native non-Clifford operation required by the BB routing circuit, so a magic-state-based implementation (requires high-fidelity magic state preparation and consumption) adds high overhead to the routing circuit. A useful QRAM protection scheme should instead exploit the dominant hardware noise and the restricted set of states visited during a valid query.

Here we develop such an architecture-specific framework under biased and erasure-dominant noise. The first component is a one-fault-tolerant logical \texttt{CSWAP} for the \([[4,2,2]]\) code. Inspired by pieceable fault tolerance~\cite{yoder2016universal}, the construction interleaves physical gates with erasure checks. When a check reports an erasure, a local recovery either undoes the executed gates and restarts the logical gate, giving a repeat-until-success (RUS) protocol, or executes the remaining gates to complete the logical operation. Consequently, no accepted execution containing one flagged erasure produces a logical error, and the leading accepted failure probability is quadratic in the physical error rate \(p_{\mathrm{phys}}\). The construction avoids magic states and is tailored to the non-Clifford routing gate that dominates BB QRAM under the erasure channel.

The second component is a QRAM-specific error-correcting code (QRAMQEC) built from \emph{intrinsic redundancy}. A valid BB query occupies only a restricted subspace of all router configurations: inactive routers remain in a wait state, while activated routers form consistent paths through adjacent tree layers. We translate such local validity constraints into parity checks on the router states. Under a bit-flip model, overlapping checks produce a decoding graph that supports minimum-weight perfect matching (MWPM); under located erasures, related constraints enable local recovery at internal routers. The routing constraints provide code redundancy without additional data qubits, although syndrome ancillas and a constant-factor router encoding are still required.

\begin{figure}[t]
    \centering
    \includegraphics[pagebox=mediabox,width=0.98\linewidth]{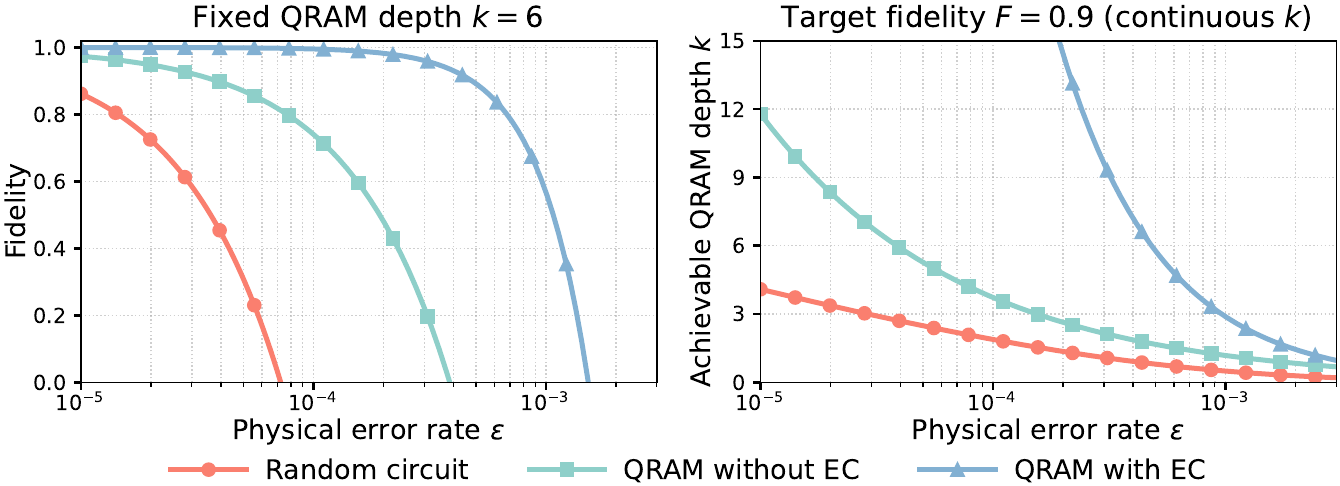}
    \caption{Illustrative fidelity models, with \(\epsilon=p_{\mathrm{phys}}\) and address width \(n\) labelled \(k\) in this figure. The left panel fixes \(n=6\) and uses \(1-F=36n2^n\epsilon\), \(72n^2\epsilon\), and \(12000n^2\epsilon^2\) for the generic circuit, uncorrected QRAM, and quadratic suppression models, respectively. The right panel shows the address width at \(F=0.9\), treated as a continuous variable, with coefficient \(144\) for the generic circuit and the same coefficients as in the left panel for QRAM. The quadratic model illustrates suppression of the depth-dependent contribution~\cite{hann}.}
    \label{fig:motivation}
\end{figure}

The two codes can be concatenated. The logical \texttt{CSWAP} provides second-order erasure protection at the inner level, while the QRAM code can additionally correct residual logical bit flips in the interior router configuration. Correcting single interior bit flips removes their leading contribution to the router-state infidelity. Combined with BB QRAM's intrinsic noise resilience~\cite{hann}, this suppression increases the memory size compatible with a fixed fidelity target, as illustrated in Fig.~\ref{fig:motivation}.

Our main contributions are as follows:
\begin{itemize}
    \item We construct a RUS logical \texttt{CSWAP} gate for the \([[4,2,2]]\) code under erasure noise. Interleaved checks and conditional recovery correct single erasures before accepting the output.
    \item We formulate a QRAM code from the restricted valid-state subspace of BB QRAM routing. The construction supports local checks and active correction under bit-flip and erasure noise without adding data qubits solely for redundancy.
    \item We evaluate the resulting protection with analytical bounds and classical simulations. The protected \texttt{CSWAP} exhibits second-order logical failure, while QRAMQEC suppresses the depth-dependent contribution of interior errors to router-state snapshot infidelity during the query from first to second order.
\end{itemize}

The remainder of the paper is organized as follows. Section~\ref{sec:background} reviews the noise models, the \([[4,2,2]]\) code, and BB QRAM. Section~\ref{sec:logicalgates} presents the protected logical \texttt{CSWAP} for the \([[4,2,2]]\) code, and Sec.~\ref{sec:qramcode} develops the QRAM code. Section~\ref{sec:results} reports the simulations and router-state snapshot scaling analysis. Section~\ref{sec:resources} quantifies the resource costs, and Sec.~\ref{sec:discussion} describes the residual noise and error scaling of the concatenated scheme.

\section{Background}
\label{sec:background}

\subsection{Noise models and a four-qubit erasure code}
Quantum noise can be described by a quantum channel, namely a completely positive trace-preserving map. In Kraus operator form,
\begin{equation}
    \mathcal{E}(\rho)=\sum_a E_a\rho E_a^\dagger,
    \qquad
    \sum_a E_a^\dagger E_a=I,
\end{equation}
where \(\rho\) is the input state and the \(E_a\) are linear Kraus
operators acting on the system Hilbert space~\cite{nielsen2002quantum}. Different physical noise processes correspond to different choices of the operators \(E_a\). In this work, we focus on two forms of structured noise: strongly biased Pauli noise, whose statistics identify a dominant error component, and erasures, whose flags identify individual fault locations.

A simple biased Pauli channel distinguishes bit- and phase-flip probabilities:
\begin{equation}
    \mathcal{E}_{\mathrm{bias}}(\rho)
    =(1-p)\rho+p_X X\rho X+p_Z Z\rho Z,
    \qquad p=p_X+p_Z .
\end{equation}
Biased noise allows codes and decoders to be tailored to the dominant Pauli component~\cite{grimm2020stabilization,tuckett2018tailoring,bonilla2021xzzx}. For the QRAM code below, we take this component to be \(X\), with bias \(\eta_X=p_X/p_Z\), and analyze the \(p_Z=0\) limit.

A different form of structure arises when the location of a fault is known. This is captured by the erasure channel. Introducing a flag state \(\ket{e}\) orthogonal to the qubit subspace, the channel is
\begin{equation}
    \mathcal{E}_{\mathrm{erase}}(\rho)
    =(1-p_e)\rho
    +p_e\,\operatorname{Tr}(\rho)\ket{e}\!\bra{e}.
\end{equation}
Located errors are generally easier to correct than unlocated errors because the decoder does not need to determine where the fault occurred~\cite{grassl1997codes}. Erasure models are therefore particularly natural for loss-dominated systems in which loss can be heralded~\cite{knill2001scheme}. Erasure conversion has been proposed for Rydberg-atom arrays and superconducting circuits~\cite{wu2022erasure,kubica2023erasure}, and erasure detection has been demonstrated for a dual-rail cavity qubit~\cite{koottandavida2024erasure}. For the concatenated construction, we use the state-selective transition \(\ket{+}\to\ket e\), whose no-flag evolution generates an \(X\)-type residual. Its Kraus representation is given in Sec.~\ref{subsec:compose_interface}.

Quantum error-correcting codes can suppress such errors by encoding logical information into a larger physical Hilbert space~\cite{shor1995scheme,steane1996multiple}. In general, let \(\{E_a\}\) be a set of linear error operators whose span
contains the faults to be corrected. A code projector \(P\) corrects this
operator set when the Knill--Laflamme condition~\cite{knill1997theory}
\begin{equation}
    P E_a^\dagger E_b P=c_{ab}P
\end{equation}
holds for every pair \(E_a,E_b\) in the set. Here \([c_{ab}]\) is a Hermitian
matrix of complex coefficients that is independent of the encoded state.

For the erasure setting considered in this work, a distance-two quantum error-correcting code is sufficient to correct a single erasure when its location is known, while also detecting an arbitrary single-qubit error. The \([[4,2,2]]\) code is the
smallest stabilizer code with this property while encoding two logical qubits
into four physical qubits. Its stabilizer generators are~\cite{gottesman1997stabilizer}
\begin{equation}
    S_Z=Z_1Z_2Z_3Z_4,
    \qquad
    S_X=X_1X_2X_3X_4 .
\end{equation}
One convenient choice of logical operators is
\begin{equation}
    \bar X_1=X_1X_3,\quad
    \bar Z_1=Z_1Z_2,\quad
    \bar X_2=X_1X_2,\quad
    \bar Z_2=Z_1Z_3 .
\end{equation}
Because the location of the erasure is known, the information contained in the remaining three physical qubits is sufficient to reconstruct both encoded logical qubits.

\subsection{Fault-tolerant logical operations}

Encoding quantum information, by itself, does not make a computation fault tolerant. Logical gates and syndrome extraction must also be designed so that a single physical fault does not spread into an uncorrectable error. Transversal gates provide a particularly simple way to achieve this: physical qubits are partitioned into independent subsets, and the logical operation acts separately on each subset, with
\begin{equation}
    U_L=\bigotimes_j U_j.
\end{equation}
Because errors cannot propagate freely between these subsets, transversal gates are naturally fault tolerant. However, no quantum code admits a universal set of transversal logical gates~\cite{eastin2009restrictions}. A fault-tolerant architecture therefore requires additional mechanisms to complete a universal gate set. Common approaches include magic-state injection and distillation~\cite{bravyi2005universal,litinski2019game}, code switching~\cite{anderson2014conversion}, and pieceable fault tolerance~\cite{yoder2016universal}.

Pieceable fault tolerance is particularly relevant to our construction. Rather than implementing a logical gate in a single fault-tolerant step, it decomposes the operation into shorter segments separated by intermediate error-correction steps,
\begin{equation}
    U_L=A_m U_m A_{m-1}\cdots A_1 U_1,
\end{equation}
where the \(U_i\) implement successive pieces of the desired logical operation, while the \(A_i\) detect or correct errors before they can accumulate across multiple pieces. Our protected \texttt{CSWAP} follows the same general principle, but specializes the intermediate operations to erasure checks and conditional recovery, together with RUS control flow.

\subsection{Bucket-brigade quantum random-access memory}
A quantum random-access memory (QRAM) implements a coherent query to classical memory by accessing multiple addresses in superposition. Specifically, it realizes the unitary operation
\begin{equation}
\sum_{i=0}^{N-1}\alpha_i\ket{i}_{\mathrm A}\ket{0}_{\mathrm B}
\xrightarrow{\mathrm{QRAM}}
\sum_{i=0}^{N-1}\alpha_i\ket{i}_{\mathrm A}\ket{x_i}_{\mathrm B},
\label{eq:qram}
\end{equation}
where \(x_i\) is the classical data stored at address \(i\), \(\alpha_i\) is the corresponding address amplitude, and the subscripts \(\mathrm A\) and \(\mathrm B\) denote the address and bus registers, respectively.

\begin{figure*}[t]
\centering
\includegraphics[width=\linewidth]{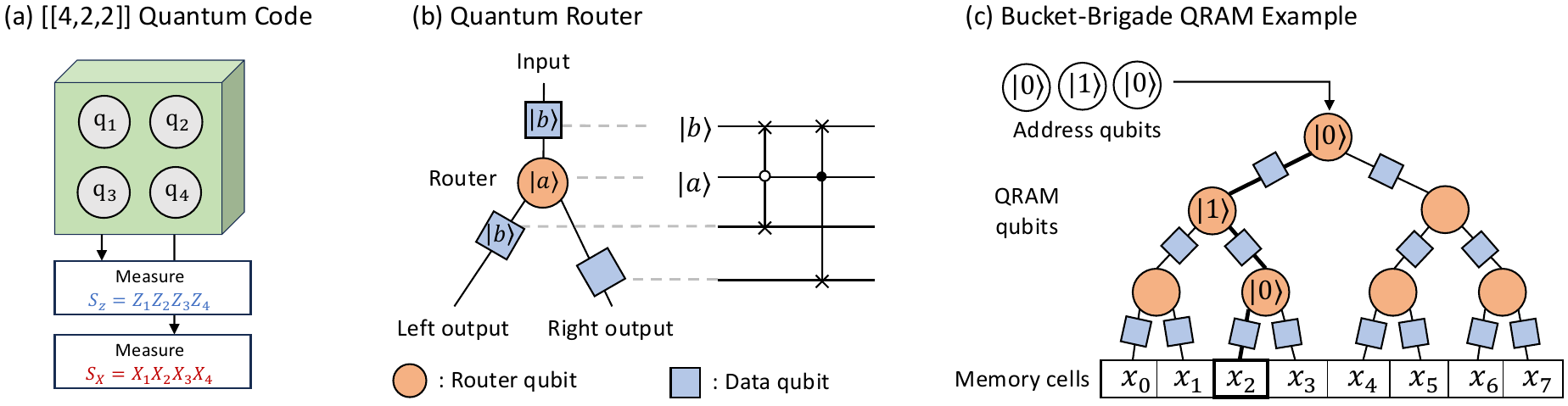}
\caption{Building blocks used in this work. (a) The \([[4,2,2]]\) erasure code encodes two logical qubits into four physical qubits; stabilizer measurements diagnose a located erasure before recovery. (b) A quantum router directs an input to one of two outputs according to its routing state. (c) Arranging routers in a binary tree yields a bucket-brigade QRAM that supports coherent queries over many addresses. In each address branch, the routers along the selected root-to-leaf path direct the incoming qubit, while the remaining routers stay in the ``wait'' state \(\ket W\). }\label{fig:background_overview}
\end{figure*}

We consider the bucket-brigade (BB) QRAM architecture proposed by Giovannetti et al.~\cite{giovannetti2008architectures,giovannetti2008quantum}. Its basic building block is a quantum router, illustrated in Fig.~\ref{fig:background_overview}(b). Each router has an inactive ``wait'' state \(\ket W\) and two active routing states that direct an incoming qubit to either its left or right output. BB QRAM recursively arranges these routers into a binary tree, as shown in Fig.~\ref{fig:background_overview}(c). A memory containing \(N=2^n\) cells has \(n\) router levels and \(N-1\) routers. We number the root level by zero and write \(k=n-1\) for the depth of the lowest router level. Each router at level \(k\) connects to two memory cells.

A BB query proceeds in three stages: \emph{address loading}, \emph{data retrieval}, and \emph{address unloading}. During address loading, the address qubits are successively routed into the tree and stored in routers at different levels. Each stored address bit controls the routing of subsequent qubits, so that a superposition of addresses coherently programs a corresponding superposition of root-to-leaf paths. In each address branch, the routers along the selected path participate in routing, while the remaining routers stay in the inactive state \(\ket W\).

Once the address has been loaded, the bus qubit is routed along the paths selected by the address to the leaves of the tree, where it interacts with the corresponding classical memory cells. The resulting data is thereby coherently written into the bus register. The bus is then routed back out of the tree, after which the address-loading procedure is reversed to uncompute the routing information and return the routers to their initial \(\ket W\) states. Because routing operations at the same tree level can be executed in parallel, the overall query depth scales as \(O(\log N)\).

Hardware proposals include hybrid quantum acoustic systems and superconducting cavities~\cite{hann2019hardware,weiss2024quantum}.

This sparse activation of the routing tree is also responsible for an important property of BB QRAM: its intrinsic noise resilience. Errors occurring in inactive routers do not directly affect the activated address branch, and the limited entanglement between different routing paths restricts error propagation. As a result, BB QRAM exhibits substantially more favorable error scaling than architectures in which a query coherently activates a macroscopic fraction of the memory~\cite{arunachalam2015robustness,hann}. In this work, the restricted set of routing configurations that can arise during a valid BB query will play an additional role: these constraints provide the intrinsic redundancy exploited by the QRAM code introduced in Sec.~\ref{sec:qramcode}.

\section{One-fault-tolerant \texttt{CSWAP} under erasure noise}
\label{sec:logicalgates}

The BB QRAM circuit considered in this work requires only two types of logical operations: single-qubit \(X\) gates and non-Clifford \texttt{CSWAP} gates. The \(X\) gates are straightforward at the encoded level, so the main missing operation is a protected \texttt{CSWAP} with sufficiently low overhead. In principle, one could realize this gate using a general-purpose non-Clifford construction, for example through magic-state-based techniques~\cite{bravyi2005universal,litinski2019game,di2020fault}. For QRAM, however, this is particularly unattractive. The router operation is repeated throughout a binary tree containing \(O(2^n)\) routers for an \(n\)-bit address space, so even a moderate overhead per logical \texttt{CSWAP} is multiplied many times across the memory.

For this reason, we construct a \texttt{CSWAP} specifically for the \([[4,2,2]]\) code and the located-erasure noise model considered here. The construction does not rely on QRAM routing constraints and also applies to logical \texttt{CSWAP} operations outside QRAM. The basic idea is closely related to pieceable fault tolerance~\cite{yoder2016universal}. Rather than executing the entire logical operation and correcting errors only at the end, we divide it into short segments and perform erasure checks between them. If an erasure is detected, the subsequent recovery depends on where in the circuit the fault occurred and on how much of the controlled swap has already taken place. The located erasure makes this construction possible with a distance-two code; the conditional recovery restores the target blocks after a partial swap.

\begin{figure}[t]
    \centering
    \includegraphics[width=\linewidth]{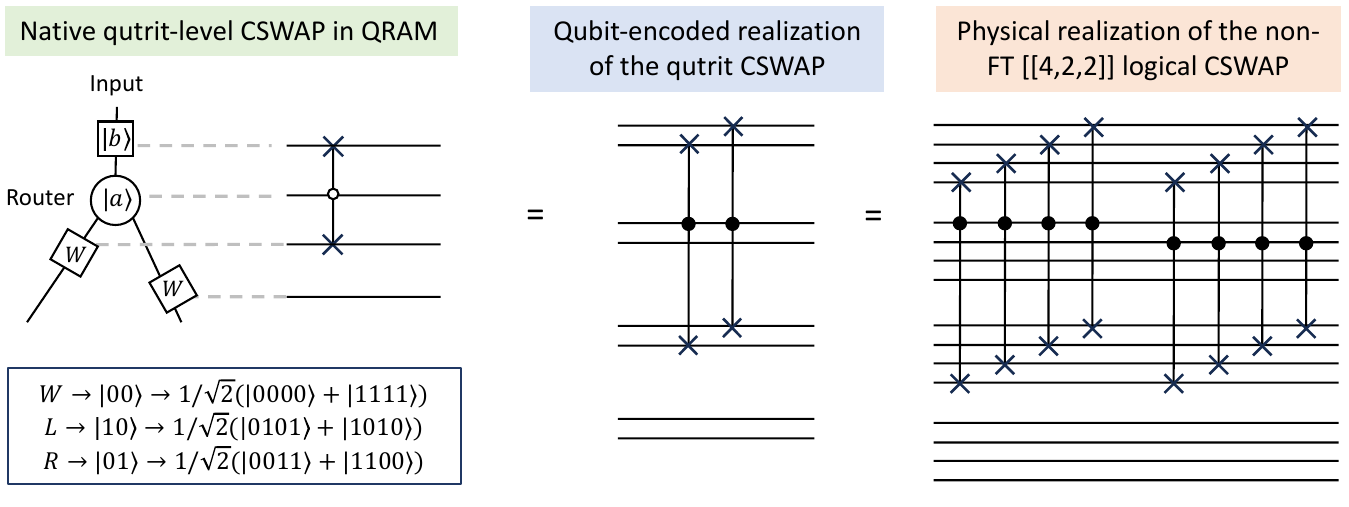}
    \caption{Interblock logical \texttt{CSWAP} constructed from eight physical \texttt{CSWAP} gates. The first logical qubit encoded in the middle block controls the exchange of the two encoded target blocks.}
    \label{fig:cswapov}
\end{figure}

\subsection{Overview}
\label{subsec:gate}

We begin with the non-fault-tolerant encoded \texttt{CSWAP} shown in Fig.~\ref{fig:cswapov}. By itself, this circuit implements the desired logical operation but is not fault-tolerant: if an erasure occurs partway through the sequence, some of the physical \texttt{CSWAP} gates may already have acted before the fault is detected. The central problem is therefore not only to repair the erased qubit, but also to determine what has happened to the two target blocks by the time the erasure is reported.

Section~\ref{subsec:correction} and Appendix~\ref{app:target-erasure-case} develop a recovery procedure that resolves this ambiguity. The basic strategy is to use the reported erasure location to decide whether to undo the executed gates or execute the remaining gates before performing local erasure correction. The protected logical \texttt{CSWAP} includes these conditional recovery operations, the erasure checks, and any RUS restarts.

\subsection{Error model and zone definitions}
\label{subsec:errormodel}

The circuit in Fig.~\ref{fig:cswapov} acts on three \([[4,2,2]]\) code blocks, containing 12 physical qubits in total. We denote the middle block by \(C\) and use it as the control block. The upper and lower blocks, denoted by \(A\) and \(B\), contain the two encoded target states. The first logical qubit encoded in \(C\) controls a blockwise swap of the two logical qubits in \(A\) with their counterparts in \(B\).

The logical operation is realized using eight physical \texttt{CSWAP} gates. It is useful to think of these gates as two consecutive groups of four. During the first half of the circuit, all four physical \texttt{CSWAP}s use the first physical qubit of block \(C\) as their control. During the second half, the same role is played by the second physical qubit of \(C\). We denote the physical control used in the current half by \(c_p\). This division into two halves will also be important for recovery. For a report in the first half, recovery undoes the executed gates and restarts the logical operation. For a report in the second half, recovery executes the remaining gates to complete the operation.

To analyze the protocol, we need to keep track of three different ways in which faults can enter the circuit. A physical qubit participating in a \texttt{CSWAP} can be erased by the gate itself; a qubit that is not participating in the current gate can suffer an erasure while it waits; and, finally, the erasure check can return the wrong classical result. We denote the corresponding elementary rates by \(\epsilon_0\), \(\epsilon_1\), and \(\epsilon_2\), respectively. It is convenient to organize these faults according to the eight physical \texttt{CSWAP}s. We therefore divide the circuit into eight \emph{steps}, one for each physical gate. A step should be understood as the full time interval associated with a physical \texttt{CSWAP}, rather than only the ideal unitary operation itself. For each of the three participating qubits, we include a gate-induced erasure location immediately before and immediately after the ideal \texttt{CSWAP}, each occurring with rate \(\epsilon_0\). During the same interval, every qubit that does not participate in the gate is subject to an idling erasure with rate \(\epsilon_1\). At the end of every step, all 12 physical qubits are checked for erasure. Each check produces a classical flag indicating whether that qubit is believed to have been erased. To model imperfect readout, we independently flip each flag with probability \(\epsilon_2\), thereby including both false positives and false negatives. In what follows, an erasure is reported at a qubit whenever its classical erasure flag is set, whether or not the underlying data qubit was actually erased.

We model the physical \texttt{CSWAP} as acting trivially on the targets when its control is erased. Cavity-controlled routing provides a physical motivation for this choice~\cite{weiss2024quantum}. For an intact control in \(\ket1\), the target states, including an erased target, are exchanged. Erasure recovery depends on whether this exchange has occurred. If the physical control is \(\ket{0}\), the gate does nothing to the two targets; if it is \(\ket{1}\), the targets are exchanged. An erasure on the control line can therefore leave an ambiguity about whether the targets have been exchanged by the executed gates. Target erasures have a related complication. Suppose, for example, that one of the target qubits is erased and a physical \texttt{CSWAP} subsequently acts on that pair. The erased degree of freedom can then be coherently exchanged between blocks \(A\) and \(B\). Before a measurement or recovery step resolves this ambiguity, it is therefore not always meaningful to assign the erasure to a unique target block. This is the main reason that recovery cannot simply consist of ``detect an erased qubit and replace it.'' By the time the fault is reported, we must also determine how the partially executed controlled swap should be interpreted.

We classify the first reported erasure using two pieces of information. The first is \emph{when} the report occurs: during the first four physical \texttt{CSWAP}s or during the final four. The second is \emph{where} the qubit reported as erased is located: in the control block \(C\), or in one of the two target blocks \(A\) and \(B\). Combining these two distinctions gives four \emph{zones}.

Zones \(1\) and \(2\) correspond to reports during the first half of the circuit, while Zones \(3\) and \(4\) correspond to reports during the second half. Within each half, an odd-numbered zone means that the qubit reported as erased belongs to \(C\), whereas an even-numbered zone means that it belongs to one of the target blocks.

\begin{figure}[t]
    \centering
    \includegraphics[width=0.8\linewidth]{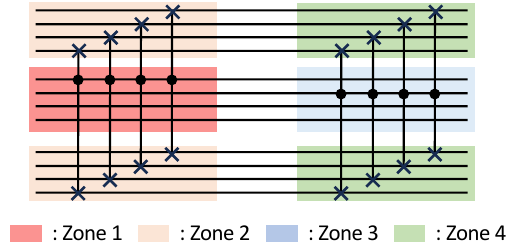}
    \caption{Possible erasure locations during the logical \texttt{CSWAP}, grouped into four recovery zones. Zones 1 and 2 correspond to reports in the first half of the circuit, while Zones 3 and 4 correspond to reports in the second half. Odd zones refer to the control block and even zones to the target blocks.}
    \label{fig:zonedef}
\end{figure}

\begin{figure}[t]
    \centering
    \includegraphics[width=\linewidth]{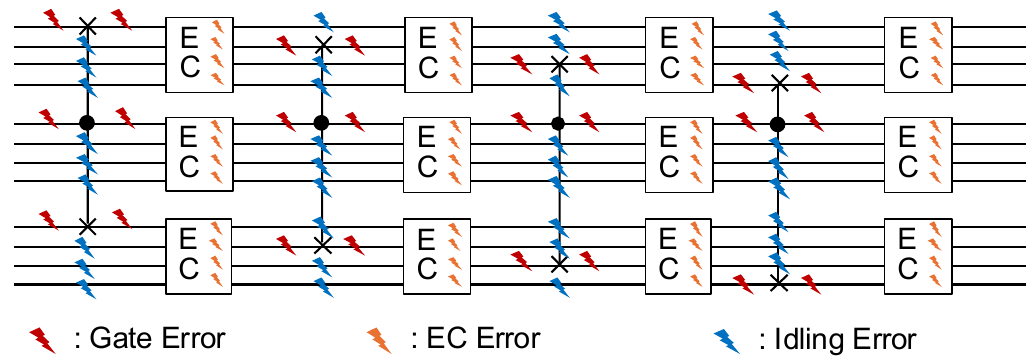}
    \caption{Error locations associated with one physical \texttt{CSWAP} step, including gate-induced erasures, idling erasures, and erasure-check errors.}
    \label{fig:errormodel}
\end{figure}

\subsection{Correction protocol}
\label{subsec:correction}

When an erasure is detected, we first undo the executed swaps or execute the remaining swaps, using the intact control qubit or the measured control value as required. Depending on when the erasure is detected, this restores the input of the logical \texttt{CSWAP} or completes its intended output. We then perform ordinary \([[4,2,2]]\) erasure recovery within the affected blocks.

\begin{figure}[t]
    \centering
    \includegraphics[width=\linewidth]{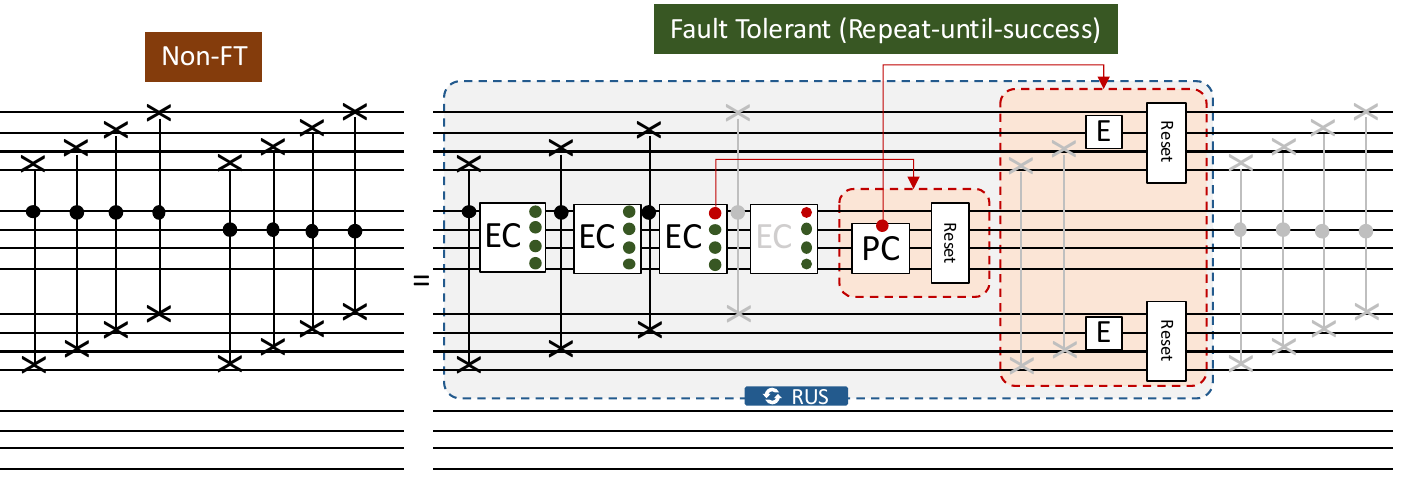}
    \caption{Overview of the repeat-until-success (RUS) correction scheme for erasures detected during the first half of the logical \texttt{CSWAP}. The executed gates are first undone, after which the local erasure is corrected and the logical gate is restarted. Reset denotes local \([[4,2,2]]\) erasure recovery. The recovery procedures for the control and target blocks are given in Fig.~\ref{fig:control-erasure-correction} and Appendix~\ref{app:target-erasure-case}, respectively.}
    \label{fig:zone1}
\end{figure}

\begin{figure*}[t]
\centering
   \includegraphics[width=0.95\textwidth]{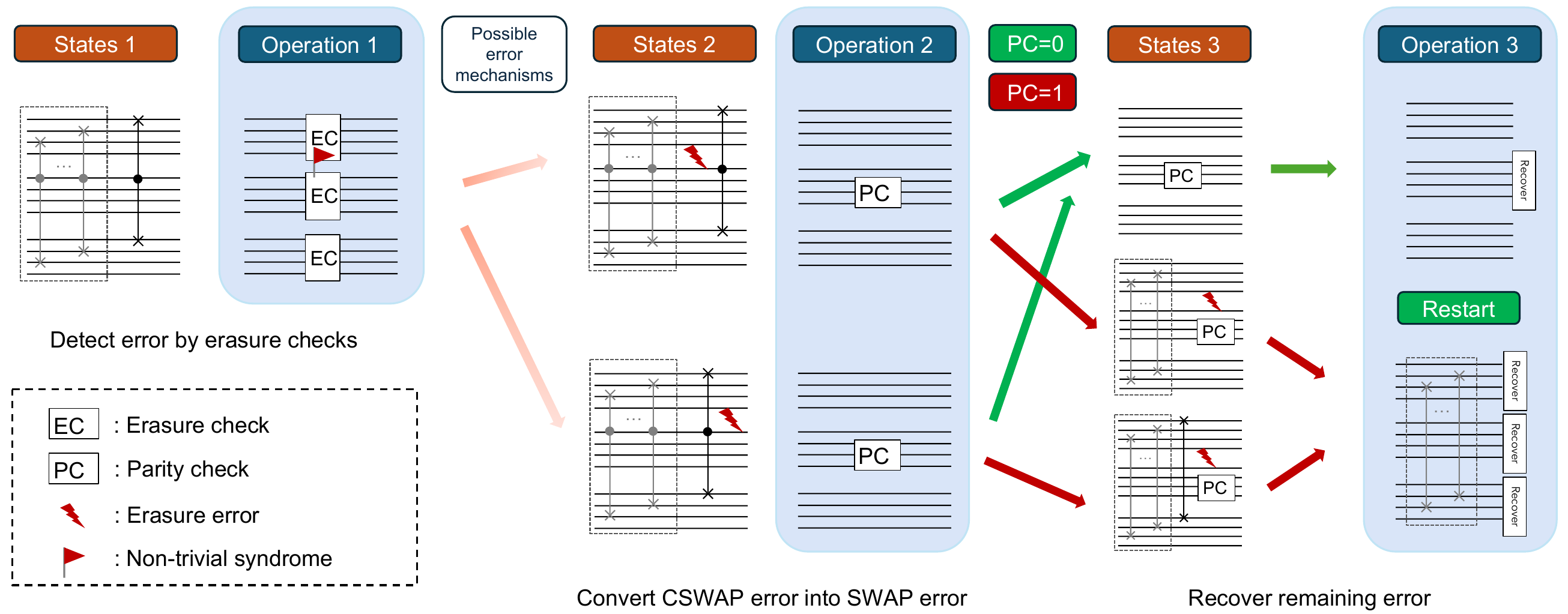}
\caption{Detailed correction flow for a reported erasure in the control block during the first half of the logical \texttt{CSWAP} gate (Zone 1). Operations 1--3 detect the erasure, obtain the measured control value (PC=0 or 1), and undo the executed swaps as needed before local recovery and restart. States 1--3 show the state of the circuit before detection, after the erasure report, and after obtaining the measured control value, respectively.}
\label{fig:control-erasure-correction}
\end{figure*}

Figure~\ref{fig:zone1} summarizes the repeat-until-success (RUS) flow for first-half reports, and Fig.~\ref{fig:control-erasure-correction} shows the detailed control-block recovery for Zone 1. The corresponding target-block recovery is given in Appendix~\ref{app:target-erasure-case}.

\subsubsection{Control-block erasures: Zones 1 and 3}

When erasure checks report an erasure in the control block \(C\) during the first half (Zone 1) or second half (Zone 3), the primary risk is that \(C\) has left the code space while targets \(A\) and \(B\) may already have undergone part of the logical swap.
Simply correcting the erased qubit in \(C\) would therefore not be sufficient. The key information is the value of the physical control used in the current half of the circuit. If that control corresponds to \(0\), the preceding \texttt{CSWAP} gates act trivially on the targets. If it corresponds to \(1\), those gates act as ordinary swaps on the target pairs. We therefore first obtain the measured control value \(b\) of the active control qubit \(c_p\). For a first-half report (Zone 1), we undo the executed swaps and later restart the gate. For a second-half report (Zone 3), we execute the remaining swaps to complete the intended output.

Suppose first that the qubit reported as erased is the active control \(c_p\) itself. Because its erasure location is known, the other three physical qubits in \(C\) still contain enough information to determine the control value. We obtain the measured control value \(b\) by measuring the joint \(Z\) parity of these three remaining qubits. The measurement supplies the physical control value needed for target-block recovery. The encoded logical information is preserved by the \([[4,2,2]]\) recovery.

The measured control value \(b\) tells us what the already executed physical \texttt{CSWAP}s have done to the target blocks:
\begin{itemize}
    \item If \(b=0\), each physical \texttt{CSWAP} acts as the identity. In this case, even though several gates may have been executed before the erasure was reported, none of them has changed the logical state of \(A\) and \(B\). The target blocks are therefore already in the input state, and no additional swap operation is required.

    \item If \(b=1\), every physical \texttt{CSWAP} executed before the report should now be interpreted as an ordinary \texttt{SWAP} on its associated target pair. The target blocks are therefore partway through the desired exchange. For a first-half report (Zone 1), we execute physical \texttt{SWAP}s on the target pairs that have already been exchanged, thereby undoing those swaps. For a second-half report (Zone 3), we execute physical \texttt{SWAP}s on the target pairs whose corresponding \texttt{CSWAP}s have not yet been executed, thereby completing the intended logical operation.

\end{itemize}

When \(b=1\) and the erased qubit is \(c_p\) itself, recovery must account for erasures both immediately before and immediately after the ideal physical \texttt{CSWAP}. If the erasure happened before the physical \texttt{CSWAP}, the current target pair \((A[i],B[i])\) has not yet been exchanged. If it happened afterward, that pair has already been swapped. Thus the two histories differ by precisely one physical \texttt{SWAP} on the current pair.

After the swaps on the unambiguous target pairs have been undone or completed, we deliberately erase both qubits in the current target pair and recover them using the \([[4,2,2]]\) correction procedures for blocks \(A\) and \(B\). Once these two located erasures are corrected independently inside the two blocks, the recovered logical state no longer depends on whether the original control erasure occurred immediately before or immediately after the physical \texttt{CSWAP}.

At this point, the target blocks have been restored to the input state or brought to the intended output state, so the ambiguity created by the controlled operation has been removed. We can then treat the remaining fault in \(C\) as an ordinary located erasure and recover the erased control-block qubit using the standard \([[4,2,2]]\) correction cycle.

On the other hand, the recovery is simpler when the qubit reported as erased belongs to \(C\) but is not the active control \(c_p\). In this case the fault is an idling erasure: the physical qubit that actually controls the current group of \texttt{CSWAP}s remains intact. We therefore retain the control information needed to manipulate the partial gate directly, without measuring the control value.

For a first-half report (Zone 1), our strategy is to return to the beginning of the logical operation. We therefore reapply the physical \texttt{CSWAP}s that have already been executed. Since each of these \texttt{CSWAP} gates is executed twice with the same intact control, their net action cancels, and the target blocks return to the input state.

For a second-half report (Zone 3), undoing the executed gates would be unnecessarily expensive because most of the logical operation has already been completed. Instead, we execute the remaining physical \texttt{CSWAP}s to complete the intended output. After these gates, the idling erasure in \(C\) is repaired using the standard \([[4,2,2]]\) correction cycle.

After a Zone-1 recovery, the protocol restarts the logical \texttt{CSWAP} from step 0. After a Zone-3 recovery, the intended logical operation is complete at step 8.

The target-block cases, Zones 2 and 4, follow the same overall logic, although the detailed recovery is different. An earlier physical \texttt{CSWAP} may already have transferred the erased qubit between \(A\) and \(B\). Appendix~\ref{app:target-erasure-case} gives the full control flow for these target-block cases.

\subsubsection{Erasure check errors}

The erasure checks introduce another possible source of error. Unlike a gate or idling erasure, a readout error does not directly modify the quantum state. Instead, it changes the classical information used by the recovery controller. It is useful to consider false positives and false negatives separately, because their effects are quite different.

A false positive occurs when the check reports an erasure even though the physical qubit is intact. Such an event may send the controller down a recovery branch that was not actually necessary, but it does not by itself damage the encoded information. The protocol can therefore respond conservatively and simply treat every reported erasure as genuine. For example, the flagged qubit can be deliberately erased and then recovered using the usual \([[4,2,2]]\) correction procedure. The cost of a false positive is consequently an increase in overhead rather than an immediate logical error. It can introduce extra recovery steps and, if it occurs during the first half of the gate, may cause an unnecessary RUS restart.

A false negative is more dangerous because a genuine erasure is present but is not reported. In that case the faulty qubit can remain in the circuit for another step before recovery is triggered. Importantly, however, a false negative becomes harmful only if two elementary events occur together: a physical erasure must first take place, and the corresponding erasure-check bit must then be flipped. The resulting contribution is therefore already second order in the elementary error rates. It scales as \(O(\epsilon_0\epsilon_2)\) when the underlying erasure is gate induced and as \(O(\epsilon_1\epsilon_2)\) when it is caused by idling.                                                                                                                                                                                     

\subsection{Logical-error analysis}
\label{subsec:logicalerror}

We compare the raw physical \texttt{CSWAP}, the non-fault-tolerant encoded implementation, and the protected logical gate. The analysis accounts for the additional erasure checks, recovery operations, and RUS restarts of the protected circuit. Detailed derivations are given in Appendix~\ref{app:cswap-derivations}.

\paragraph{Baselines.}

A single physical \texttt{CSWAP} acts on three qubits. Under the before-and-after convention used in our error model, each of the three participating qubits has one gate-erasure location before the ideal gate and one afterward. There are therefore six first-order gate-erasure locations in total. Since all three qubits participate in the gate, there are no idling locations in this reference circuit. The resulting state-independent fidelity is
\begin{align*}
    F = (1-\epsilon_0)^6
    \approx 1 - 6\epsilon_0.
\end{align*}

The non-fault-tolerant encoded \texttt{CSWAP} in Fig.~\ref{fig:cswapov} contains eight such physical gates and also exposes the remaining qubits to idling noise while each gate is applied. Counting all of these first-order locations gives
\begin{align*}
    F
    \leq
    (1-\epsilon_0)^{48}(1-\epsilon_1)^{72}
    \approx
    1 - 48\epsilon_0 - 72\epsilon_1.
\end{align*}
Thus, before any fault-tolerant recovery is added, the encoded implementation is actually more vulnerable than a single raw physical \texttt{CSWAP} simply because it contains many more physical locations at which an erasure can occur.

The coefficients appearing here come directly from counting first-order fault locations; they are not obtained by fitting simulation data. If, for the purpose of comparison, we identify the gate and idle erasure rates with a common physical rate \(p_{\mathrm{phys}}\), the two reference error probabilities become
\(
6p_{\mathrm{phys}}
\)
for a single raw physical \texttt{CSWAP}, and
\(
120p_{\mathrm{phys}}
\)
for the non-fault-tolerant encoded implementation. These linear reference lines provide the baselines against which the quadratic scaling of the protected construction should be compared.

\paragraph{Number of attempts.}

The first half of the protected circuit is not necessarily executed only once. Whenever a first-half report occurs (Zone 1 or 2), the executed gates are undone and the logical \texttt{CSWAP} is attempted again. To leading order, the probability that one first-half attempt completes without triggering such a restart is
\(
1-24\epsilon_0-36\epsilon_1-48\epsilon_2.
\)
The number of attempts follows a geometric distribution, giving
\begin{align*}
\mathbb{E}[\text{attempts}]
\approx
\frac{1}{
1-24\epsilon_0-36\epsilon_1-48\epsilon_2
}.
\end{align*}

Not every unsuccessful attempt executes all four first-half physical \texttt{CSWAP}s before the erasure is reported. Accounting for the point at which the report occurs gives the expected number of physical \texttt{CSWAP} steps spent in the repeatable first half,
\begin{align*}
\mathbb{E}[\text{steps}_{\mathrm{first}}]
\approx
\frac{
4-36\epsilon_0-54\epsilon_1-72\epsilon_2
}{
1-24\epsilon_0-36\epsilon_1-48\epsilon_2
}.
\end{align*}

The second half does not have the same repeat structure: if a report occurs there, the recovery completes the intended output rather than restarting the gate. Conservatively counting all four physical gates in the second half gives the upper estimate
\[
4+\mathbb{E}[\text{steps}_{\mathrm{first}}].
\]
In particular, a second-half erasure can trigger recovery that completes the swap before all four remaining physical \texttt{CSWAP}s are executed, so some trajectories terminate with fewer gate steps than counted here.

\paragraph{Fault-tolerant logical error rate.}
The purpose of the recovery protocol is to correct single faults before accepting the output. Once those single-fault errors have been corrected, the leading residual failures necessarily involve pairs of elementary faults. The fault-pair classification derived in Appendix~\ref{app:cswap-derivations} gives the following second-order estimate:
\begin{align*}
p_{\textrm{FT}} \approx {}&
96\epsilon_0^2 + 72\epsilon_1^2 +360\epsilon_2^2 \\
&+ 744\epsilon_0\epsilon_1
+ 192\epsilon_0\epsilon_2
+ 288\epsilon_1\epsilon_2 .
\end{align*}
Each term corresponds to a different combination of gate, idling, and erasure-check faults that can evade the single-fault recovery logic.

For comparison, the corresponding leading error probabilities of the two unprotected reference circuits are
\begin{align*}
    p_{\textrm{non-FT}}
    &\approx 48\epsilon_0 + 72\epsilon_1,\\
    p_{\textrm{raw}}
    &\approx 6\epsilon_0.
\end{align*}
Both are linear in the elementary physical error rates, in contrast to the quadratic expression obtained for the protected logical gate.

Erasure checks, conditional recovery, and RUS restarts add physical fault locations. The resulting tradeoff is particularly clear for the parameter choice
\(
\epsilon_0=\epsilon_2=10\epsilon_1.
\)
At relatively large physical error rates, the sizable second-order prefactors in \(p_{\textrm{FT}}\) can outweigh the benefit of fault tolerance, and the non-fault-tolerant encoded implementation can perform better. The break-even point is therefore set by the competition between the smaller linear coefficients of the non-fault-tolerant encoded circuit and the larger but second-order coefficients of the protected one. The protected logical \texttt{CSWAP} changes the leading failure probability from first to second order.

\subsection{Limiting the number of RUS attempts}
The RUS expressions above formally allow the first half of the logical gate to be retried an arbitrary number of times. Such an unbounded loop is useful analytically, but it is not necessary if our goal is only one-fault tolerance.

A third attempt requires erasure reports in the first half of two successive attempts, and therefore at least two faults. Limiting the gate to two attempts preserves one-fault tolerance and bounds the number of executions of the original eight physical \texttt{CSWAP}s, including restarts. The controller reports an abort when a third attempt would be required.

Let
\(
r =
24\epsilon_0+36\epsilon_1+48\epsilon_2
\)
denote the leading probability that a single first-half attempt triggers a restart. Requiring a third attempt means that this has happened twice. Thus, considered separately from other detected failures, the probability that no third attempt is required is
\begin{align*}
P_{\textrm{no third attempt}}
&=
1-r^2 \\
&=
1-\left(
24\epsilon_0+36\epsilon_1+48\epsilon_2
\right)^2.
\end{align*}

The leading fault-pair estimate for the protected circuit remains
\begin{align*}
p_{\textrm{FT}}
\approx {}&
96\epsilon_0^2
+72\epsilon_1^2
+360\epsilon_2^2 \\
&+744\epsilon_0\epsilon_1
+192\epsilon_0\epsilon_2
+288\epsilon_1\epsilon_2 .
\end{align*}

It is important not to treat every non-successful trajectory as a deterministic logical failure. There are at least three distinct outcomes. First, the controller may terminate because a third attempt would be required. Second, another fault pattern may be detected explicitly and cause the protocol to abort. Third, the protocol may accept the trajectory but nevertheless produce an incorrect logical output. Only the last case causes an undetected logical error of an accepted execution.

\section{QRAM code from routing constraints}
\label{sec:qramcode}

\begin{figure*}[t]
    \centering
    \includegraphics[width=1.0\linewidth]{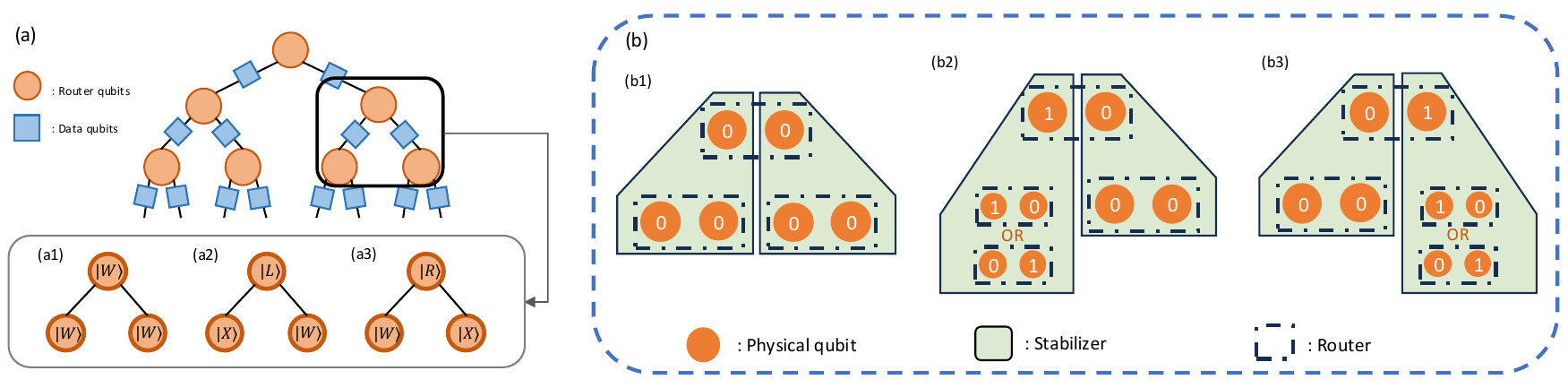}
    \caption{
    (a) Local validity constraints imposed by bucket-brigade (BB) routing. Each router qutrit takes values in \(\{\ket{W},\ket{L},\ket{R}\}\), with \(\ket{X}\in\{\ket{L},\ket{R}\}\) denoting an activated router.
    (b) Dual-rail representation of the local constraints. Each green region supports a \(Z\)-type parity check.
    }
    \label{fig:static}
\end{figure*}

The routing constraints of BB QRAM provide redundancy for error correction. During an ideal query, the router register cannot explore an arbitrary state of the full QRAM Hilbert space. Instead, the routing rules restrict it to a much smaller set of configurations in which the routers along an active branch are mutually consistent. These constraints are normally viewed simply as part of the operation of the QRAM. Here, we instead treat them as information that can be checked for errors.

These local routing constraints define a code space, and violations of those constraints provide syndrome information. Because the constraints involve only neighboring routers, they can be tested using sparse local checks. A nontrivial syndrome can then be used either to detect an error or, together with the decoder of Sec.~\ref{subsec:decoder}, to actively correct it during the QRAM query.

The correlations used for protection are already present in a valid BB routing state. The implementation uses local syndrome ancillas and a constant-size physical encoding of each router. We develop this construction for bit-flip and erasure noise in Secs.~\ref{subsec:bitflip} and~\ref{subsec:erasure}, respectively.

Consider an \(n=k+1\)-level BB tree in which each router is represented by a qutrit with computational-basis states
\begin{equation}
    \ket{W}\ \text{(wait)},\qquad
    \ket{L}\ \text{(route left)},\qquad
    \ket{R}\ \text{(route right)}.
\end{equation}

After the address has been loaded into the tree, its information is also encoded in the pattern of activated routers across different layers. For any basis-state address, only the routers along the corresponding path are activated; routers away from that path remain in \(\ket{W}\). Moreover, the choices made at neighboring levels must be mutually consistent. For a superposition of addresses, the router register is therefore a coherent superposition of these sparse activation patterns consistent with the routing constraints.

The restriction can already be seen locally. Consider a router \(u\) above level \(k\), with child routers \(u_L\) and \(u_R\), after address loading, as shown in Fig.~\ref{fig:static}(b). Their states obey the following local constraints. If the parent router is waiting, both children are inactive. If the parent routes left, its left child is active and its right child waits; a right-routing parent has the opposite pattern.

Using \(\ket{X}\in\{\ket{L},\ket{R}\}\) to denote an activated child router, the corresponding allowed local subspace is
\begin{align}
    \mathcal{V}_u \;=\; \mathrm{span}\Big\{&
    \ket{W}_u\ket{W}_{u_L}\ket{W}_{u_R}, \nonumber\\
    &\ket{L}_u\ket{X}_{u_L}\ket{W}_{u_R},\;
      \ket{R}_u\ket{W}_{u_L}\ket{X}_{u_R}
    \Big\}.
    \label{eq:local_valid_subspace}
\end{align}

Equivalently, if the parent is in $\ket{W}$ then both children must be in $\ket{W}$; if the parent is in $\ket{L}$ (respectively $\ket{R}$) then the left (respectively right) child is activated and the other child must remain in $\ket{W}$. These implications are the \emph{intrinsic redundancy}: child activation implies parent activation, and such implications overlap across layers because multiple address bits occupy different levels in the tree, as illustrated for address \(\ket{1101}\) in Fig.~\ref{fig:bitflip}(a).

We can formalize this picture by viewing the router states that are reachable at a fixed stage of an ideal query as a code space \(C\) embedded in the much larger QRAM Hilbert space. Each internal router supplies a local validity constraint, and the simultaneous satisfaction of all such constraints defines the allowed global subspace. Because every constraint is local to a parent router and its children, these checks remain geometrically local. At the same time, neighboring checks overlap across tree layers, producing the repetition-like syndrome structures that will be used below for decoding.

The rest of this section applies this principle to two representative noise channels. Under bit-flip-biased noise, a dual-rail representation of each router turns the routing constraints into simple \(Z\)-type parity checks, and the overlap of those checks leads naturally to an MWPM decoder. Under erasure noise, we use a different router encoding and exploit the same parent-child constraints to reconstruct a locally erased qubit.

\subsection{Bit-flip noise}
\label{subsec:bitflip}

\begin{figure*}[t]
    \centering
    \includegraphics[width=1.0\linewidth]{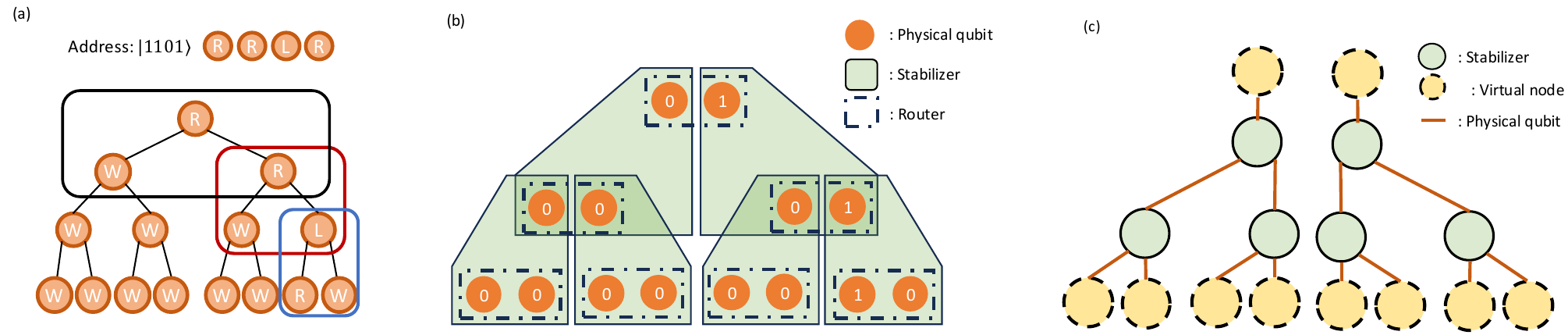}
    \caption{(a) Routing constraints for address \(\ket{1101}\) overlap across tree layers. (b) Under Pauli-\(X\) noise, these relations are measured using overlapping local \(Z\)-type checks, each supported on a parent router and its two children. (c) With the dual-rail router encoding, the resulting decoding graph separates into two tree components that can be decoded independently by MWPM.}
    \label{fig:bitflip}
\end{figure*}

We first consider a simple i.i.d.\ bit-flip model in which each physical qubit undergoes
\begin{equation}
    \rho \;\mapsto\; (1-\epsilon)\rho + \epsilon\, X\rho X,
\end{equation}
where \(\epsilon\) is the bit-flip probability per qubit during the relevant time step. This biased model isolates the type of fault that the parity checks below are designed to diagnose.

To express the routing constraints in a form that can be measured with standard two-qubit gates, we encode each router qutrit into two physical qubits, called the left and right rails, using the dual-rail mapping
\begin{equation}
    \ket{W}\mapsto \ket{00},\qquad
    \ket{L}\mapsto \ket{10},\qquad
    \ket{R}\mapsto \ket{01}.
    \label{eq:dual_rail_map}
\end{equation}
The fourth two-qubit state, \(\ket{11}\), is not used and is treated as outside the intended router subspace. This encoding has a useful feature for bit-flip detection: the routing direction is represented directly by which rail carries the excitation. As a result, the parent-child consistency conditions of Eq.~\eqref{eq:local_valid_subspace} translate into low-weight parity relations among nearby rails rather than requiring a general qutrit measurement.

Consider a single BB element consisting of one internal router and its two children. Once the address has been loaded, the valid dual-rail configurations of these three routers obey two independent parity relations. Label the parent's left and right rails by 1 and 4, the left child's rails by 2 and 3, and the right child's rails by 5 and 6. The local \(Z\)-type operators are
\begin{equation}
    \mathcal{S}_{\mathrm{elem}}
    =
    \{Z_1 Z_2 Z_3,\; Z_4 Z_5 Z_6\},
\end{equation}
or equivalently described by the parity-check matrix
\begin{equation}
    H \equiv
    \begin{bmatrix}
    1 & 1 & 1 & 0 & 0 & 0\\
    0 & 0 & 0 & 1 & 1 & 1
    \end{bmatrix}.
\end{equation}

Every valid local routing configuration lies in the \(+1\) eigenspace of these two checks. Any single-qubit error \(X_j\) anticommutes with at least one operator in \(\mathcal{S}_{\mathrm{elem}}\), and therefore changes at least one measured parity. A single bit flip is consequently detectable using only the local BB element. The two checks leave a 16-dimensional \(+1\) eigenspace containing the five-dimensional routing subspace \(\mathcal{V}_u\); they diagnose the targeted Pauli-\(X\) faults without testing all routing constraints.

One possibility would be to use these checks only for error detection and abort whenever a nontrivial syndrome is observed. However, such a strategy scales poorly for QRAM. A BB tree with \(n\) router levels contains \(O(2^n)\) routers, so at any fixed physical error rate the probability that every local check remains trivial decreases rapidly with system size. Moreover, aborting the QRAM call may require discarding the larger quantum computation in which the query is embedded.

We therefore use the syndrome for active correction rather than postselection. When the same local checks are applied throughout the tree, neighboring checks along a branch overlap on rails, as shown in Fig.~\ref{fig:bitflip}(c). Away from a boundary, a single \(X\) error then flips two neighboring checks and appears as a pair of syndrome defects. The resulting graph has the same basic structure that makes repetition-code decoding possible: error chains connect syndrome defects or terminate at a boundary, and longer chains become progressively less likely under sparse stochastic noise.

\subsection{MWPM decoding}
\label{subsec:decoder}

The overlapping parity checks naturally define a graph on which Pauli-\(X\) faults can be decoded using minimum-weight perfect matching (MWPM)~\cite{dennis2002topological,higgott2025sparse}. Let
\(\mathcal{S}=\{S_i\}\) denote the measured \(Z\)-type checks. We construct a decoding graph \(G_D=(V_D,E_D)\) by associating one vertex \(v_i\) with each check \(S_i\).

A rail \(q\) that belongs to two checks \(S_i\) and \(S_j\) is represented by an edge
\[
e(q)=\langle v_i,v_j\rangle .
\]
This graph has a direct error interpretation. A bit flip \(X_q\) anticommutes with each \(Z\)-type check containing \(q\), so if \(q\) lies between two checks, both outcomes change sign. A single interior error therefore creates a pair of defects at the endpoints of the corresponding decoding-graph edge.

The root and the leaves require special treatment because rails near these boundaries may participate in only one check. A single \(X\) error on such a rail produces only one observed defect rather than a pair. We represent this open-boundary behavior by adding virtual boundary vertices to the decoding graph. In particular, we include virtual vertices for the root region and for the leaf region, and connect nearby check vertices to the appropriate boundary. For \(k\ge2\), each root rail has a unique single-fault syndrome and is corrected by MWPM in the exact-syndrome, i.i.d.\ model. The ambiguity between the two rails of a leaf router gives the first-order boundary contribution. MWPM can then explain an observed defect in two ways: it may pair the defect with another syndrome defect, corresponding to an error chain between them, or pair it with a virtual boundary, corresponding to a chain that terminates at the edge of the code.

To make the matching sensitive to the physical noise model, each decoding-graph edge is assigned a likelihood-derived weight. If the rail associated with edge \(e(q)\) has bit-flip probability \(\epsilon_q\), we use
\begin{equation}
    w\big(e(q)\big)
    =
    -\ln\!\left(\frac{\epsilon_q}{1-\epsilon_q}\right).
\end{equation}
The cost of an error chain is then the sum of its edge weights, which is proportional to its negative log-likelihood up to an overall constant. In the i.i.d.\ model, where all \(\epsilon_q=\epsilon\), every decoding-graph edge has the same weight and the comparison between candidate error chains reduces to their path length.

Given a measured syndrome, we call every check with outcome \(-1\) a defect. These defect vertices, together with the virtual boundary vertices, define the syndrome graph \(G_S=(V_S,E_S)\). For each pair \(u,v\in V_S\), we assign an edge whose weight is the minimum path cost between \(u\) and \(v\) in the underlying decoding graph \(G_D\). MWPM then chooses a set of defect-defect and defect-boundary pairings with minimum total cost. Each matched pair determines a minimum-weight path in \(G_D\), which in turn specifies a candidate set of rails on which an \(X\) correction should be applied. For the dual-rail QRAM construction, the check structure separates at the root of the tree. The decoding graph therefore decomposes into two disjoint tree components, allowing the two sides of the QRAM tree to be decoded independently.

\subsection{Boundary-limited distance and interior protection}
\label{subsec:distance}

The resulting QRAM code should not be viewed as a conventional uniform-distance code. Its decoding graph follows the tree geometry, and the leaf boundaries are less protected than the interior. Although every single \(X\) error produces a nontrivial syndrome, that syndrome does not always identify a unique correction. Each leaf rail participates in only one check, so an error can produce a single defect that is naturally connected to the virtual leaf boundary. Moreover, errors on the two rails of the same leaf router can produce the same syndrome, as illustrated in Fig.~\ref{fig:bitflip}. This boundary ambiguity fixes the global open-boundary \(X\)-distance at
\(
d_X=2.
\)

The global distance, however, does not capture all of the protection that is relevant for QRAM operation. The code is strongly position dependent. In the decoding graph, a rail is represented by an edge, while a multi-qubit error chain is represented by a path. For a router deep in the interior of the tree, the surrounding syndrome vertices are separated from the root and leaf boundaries by many graph edges. Confusing an interior error with a chain that terminates at a boundary therefore requires a correspondingly longer pattern of physical faults. As the physical error rate \(\epsilon\) decreases, these longer chains become rapidly less likely. The situation is analogous to an open repetition code: its formal distance is set by its weakest boundary, but errors well inside the tree can nevertheless receive substantially stronger protection.

\subsection{Erasure noise}
\label{subsec:erasure}

We now turn to erasure noise, where the important additional information is that the location of the lost qubit is known. For a physical qubit we use the standard Kraus representation
\begin{align}
\mathrm{Erasure}=\big\{&
\sqrt{1-\varepsilon}\left(\ket{0}\!\bra{0}+\ket{1}\!\bra{1}\right),\nonumber\\
&\sqrt{\varepsilon}\ket{e}\!\bra{0},\;
\sqrt{\varepsilon}\ket{e}\!\bra{1}
\big\},
\end{align}
where \(\ket{e}\) is orthogonal to the computational qubit subspace and the erased location is flagged.

For erasure correction, we encode each internal router qutrit into a three-qubit block
\[
u=(u_1,u_2,u_3)
\]
according to
\begin{align}
\ket{W}_u &= \frac{1}{\sqrt{2}}\left(\ket{100}+\ket{011}\right), \nonumber\\
\ket{L}_u &= \frac{1}{\sqrt{2}}\left(\ket{001}+\ket{110}\right), \nonumber\\
\ket{R}_u &= \frac{1}{\sqrt{2}}\left(\ket{010}+\ket{101}\right).
\label{eq:3q_router_encoding}
\end{align}

Each logical router state is a coherent superposition of two complementary three-bit strings. If one physical qubit is erased, the two surviving qubits retain part of the information needed to reconstruct the original codeword. The BB routing constraint between the parent router and its children supplies the remaining information needed to identify the erasure branch. Recovery thus uses the local encoding together with the joint routing state of the tree.

The child information needed for recovery can be summarized by a simple activation parity. For any router block \(v\), define
\begin{equation}
A(v) \;=\; v_2 \oplus v_3 ,
\label{eq:activation_parity}
\end{equation}
where \(\oplus\) denotes XOR of the computational-basis values. Evaluating this quantity on the three encoded router states in Eq.~\eqref{eq:3q_router_encoding} gives
\(
A(W)=0,
A(L)=A(R)=1.
\)

Let $u$ be an internal router with children $u_L$ and $u_R$ (each itself a three-qubit block). In the ideal BB valid subspace, the parent state is determined by the child activation pattern: $(A(u_L),A(u_R))=(0,0)$ corresponds to $\ket{W}_u$, $(1,0)$ to $\ket{L}_u$, and $(0,1)$ to $\ket{R}_u$. We will only use $A(u_L)$ or $A(u_R)$ inside a syndrome bit that is constant across $\{\ket{W},\ket{L},\ket{R}\}$ for a fixed erasure branch, so measuring the syndrome does not collapse the logical superposition at $u$. For a coherent superposition of routing configurations, each erasure branch contains the corresponding superposition of surviving basis patterns. The branch depends on whether the erased qubit carried a \(0\) or a \(1\) in the corresponding component of Eq.~\eqref{eq:3q_router_encoding}. The first task of recovery is therefore to determine which of these two erasure branches occurred.

A fresh qubit is then initialized to \(\ket{0}\) at the erased location and extracts a single syndrome bit \(s\in\{0,1\}\). The syndrome is designed so that it distinguishes the two erasure branches but is otherwise insensitive to whether the logical router state is \(W\), \(L\), or \(R\). This distinction is essential: the recovery is allowed to learn how the erasure occurred, but not which route is present in the address superposition.

The required syndrome depends on which physical location was erased. Define
\begin{align}
s(u;1) &= u_2 \oplus A(u_L), \label{eq:syndrome_j1}\\
s(u;2) &= u_1 \oplus A(u_L), \label{eq:syndrome_j2}\\
s(u;3) &= u_1 \oplus A(u_R). \label{eq:syndrome_j3}
\end{align}
Here the parent bit appearing in each expression is one of the surviving physical qubits, while the child activation parity supplies the routing constraint needed to interpret it.

Consequently, measuring \(s(u;j)\) reveals enough information to perform the recovery but does not reveal whether the router at \(u\) is in \(\ket{W}\), \(\ket{L}\), or \(\ket{R}\). The joint routing state has the form
\[
\alpha\ket{W}_u\ket{\chi_W}+\beta\ket{L}_u\ket{\chi_L}
+\gamma\ket{R}_u\ket{\chi_R},
\]
where \(\ket{\chi_W},\ket{\chi_L},\ket{\chi_R}\) describe the children and the rest of the tree with the corresponding valid activation patterns. For erased computational-basis value \(b\), each syndrome in Eqs.~\eqref{eq:syndrome_j1}--\eqref{eq:syndrome_j3} equals \(1-b\), preserving coherence between these routing configurations.

Once the syndrome has identified the erasure branch, the remaining recovery consists of two simple operations:

\emph{Step 1: conditional \(X\) corrections.}
If \(s(u;j)=1\), we apply \(X\) to each of the two surviving physical qubits in the parent block. If \(s(u;j)=0\), no operation is required. This removes the difference between the two erasure branches while preserving the logical amplitudes \(\alpha,\beta,\gamma\).

\emph{Step 2: restore the parent block with a fixed Clifford circuit.}
Let \(u_j\) denote the erased location and \(u_k,u_\ell\) the two surviving qubits. After the syndrome readout, we reset the replacement qubit at \(u_j\) to \(\ket{0}\) and apply
\begin{equation}
U_{\mathrm{rec}}^{(j)}
=
X_{u_j}\;
\mathrm{CNOT}(u_j\!\rightarrow\!u_k)\;
\mathrm{CNOT}(u_j\!\rightarrow\!u_\ell)\;
H_{u_j}.
\label{eq:reencode_clifford}
\end{equation}
Because Step 1 has already removed the branch-dependent ambiguity, the same Clifford circuit can be used regardless of which erasure branch occurred. This operation restores the parent block to the encoded subspace of Eq.~\eqref{eq:3q_router_encoding} and restores its coherence with the children and the rest of the tree.

The complete recovery is therefore local and Clifford. It requires one syndrome measurement, a reset of the replacement qubit, at most two Pauli-\(X\) corrections on the surviving parent qubits, and the fixed re-encoding circuit of Eq.~\eqref{eq:reencode_clifford}. The same local rule also determines the natural recovery order when erasures occur at several internal routers. The syndrome for a damaged parent uses activation information from its children, so those child blocks must be valid before they are used for recovery. For at most one erasure per affected block, with the required child information available or recoverable, we process erasures from the bottom of the tree upward. Routers at the greatest affected depth are repaired first; once those child blocks have been restored, their parents can be corrected using the same local procedure. If only one router is affected, the protocol reduces to one local recovery.

Leaf routers are the one exception. Because a leaf has no children, the routing constraint involving the parent and its children that supplies the missing syndrome information is unavailable. Erasures at this boundary can still be detected, but the local recovery mechanism described above cannot reconstruct them. We therefore treat leaf erasures as detected but uncorrected boundary events in the present analysis.

\section{Results}
\label{sec:results}

\subsection{Simulation protocols and metrics}
\label{subsec:simulation-details}

We evaluate the two protection layers using separate classical simulations, since they operate under different noise models and require different performance metrics.

A quantum-trajectory Monte Carlo simulation examines the leading error scaling of the protected \([[4,2,2]]\) repeat-until-success (RUS) logical \texttt{CSWAP} under state-independent erasure noise. After each physical \texttt{CSWAP}, the simulator applies an effective erasure channel to the participating qubits, an idling-erasure channel to all other qubits, and then samples the erasure-check outcomes. It also follows the full classical control flow of the protected logical gate, including false-positive and false-negative reports, conditional recovery that undoes the executed gates or executes the remaining gates, ideal local erasure correction, and RUS restarts.

For the data reported here, we set
\(
\epsilon_0=\epsilon_1=\epsilon_2=p_{\mathrm{phys}},
\)
and allow at most two attempts (one RUS restart). The recorded gate-failure probability \(p_L\) includes three outcomes: an incorrect final logical state, an abort due to uncorrectable errors, or requiring a second RUS restart. Adaptive sampling with between \(200\) and \(50{,}000\) trajectories at each physical error rate is used for simulation efficiency. Specifically at \(p_{\mathrm{phys}}=10^{-4}\), no failures were observed in \(50{,}000\) trajectories, so we report the Wilson-score 95\% upper bound
\[
p_L<7.7\times10^{-5}
\]
rather than assigning a nonzero point estimate. The fit in Fig.~\ref{fig:cswap_simulation} shows suppression of the leading first-order error contribution with an estimated break-even error rate at \(p_{\mathrm{phys}}=3 \times10^{-2}\), making the scheme practical even for current noisy quantum hardware.

The raw curve represents the dual-rail routing operation, which uses two physical \texttt{CSWAP}s on six qubits. Each step has six active before/after-gate noise locations and three idle locations. The raw and non-fault-tolerant encoded baselines are therefore
\[
12\epsilon_0+6\epsilon_1
\qquad\text{and}\qquad
48\epsilon_0+72\epsilon_1,
\]
which reduce to \(18p_{\mathrm{phys}}\) and \(120p_{\mathrm{phys}}\) when all three rates equal \(p_{\mathrm{phys}}\).

QRAMQEC is evaluated with a separate \texttt{Stim}-based stabilizer simulation~\cite{gidney2021stim} under the bit-flip model of Sec.~\ref{subsec:bitflip}. For each tree depth and physical error rate, we initialize the valid all-\(\ket{W}\) router configuration, perform one noisy round of ancilla-based \(Z\)-check extraction, and finally measure all data qubits. Data-idling, ancilla-idling, post-\texttt{CNOT}, check-readout, and final-data-readout bit-flip rates are all set to \(p_{\mathrm{phys}}\). The MWPM decoder receives both the ancilla detector events and final-data parity detectors obtained by comparing the final ancilla outcomes with the corresponding parities of the data readout. The resulting correction is applied to the measured router-state snapshot as a Pauli-frame update. Because the detector response to Pauli-\(X\) faults is independent of which valid address pattern is loaded, the same residual-error statistics can be used to evaluate arbitrary valid router-state snapshots. This benchmark evaluates the decoded router-state snapshot. Each archived point in Fig.~\ref{fig:qram_tree_state_simulation} averages \(100{,}000\) shots.

We quantify QRAMQEC performance using \(F_{\mathrm{tree}}\), defined as the total probability weight of leaves whose complete root-to-leaf paths are correct after decoding. The leaf weights are uniform in our simulations. The weights of boundary edges and other decoding-graph edges are inherited from the circuit-level detector-error model, and matching degeneracies are resolved by the decoder implementation. We score the corrected routing state rather than asking whether the decoder identifies the exact fault configuration, due to the noise resilience of QRAM~\cite{hann2021practicality}, which allows every routing path to contribute individually to the output state of the query.

% For the protected \texttt{CSWAP}, error bars are Wilson-score 95\% confidence intervals. For QRAMQEC, only the aggregate means remain in the archived data, so we use the conservative bounded-variable estimate
% \[
% 1.96\sqrt{
% \frac{F_{\mathrm{tree}}(1-F_{\mathrm{tree}})}
% {n_{\mathrm{shot}}}
% },
% \qquad
% n_{\mathrm{shot}}=100{,}000,
% \]
% rather than estimating the uncertainty from unavailable per-shot samples.

\subsection{Numerical results}

\begin{figure}[t]
  \centering
  \includegraphics[width=\linewidth]{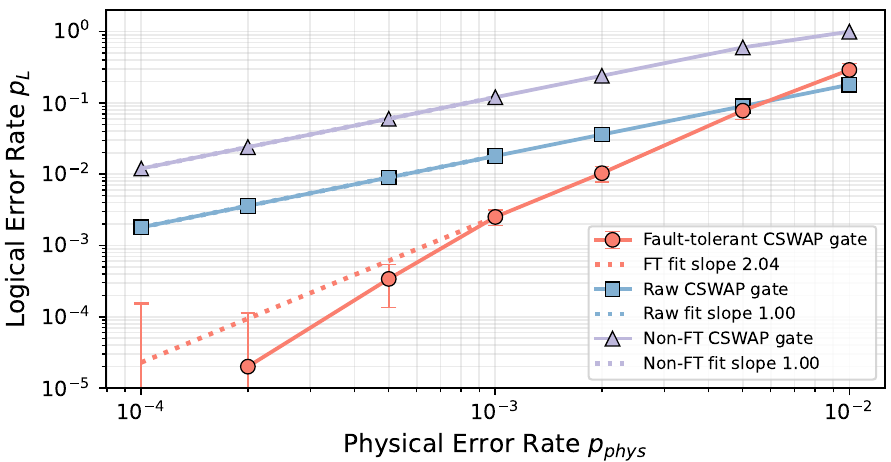}
  \caption{
  Failure probability \(p_L\) versus physical error rate \(p_{\mathrm{phys}}\) for the raw dual-rail routing operation, the non-fault-tolerant encoded construction of Fig.~\ref{fig:cswapov}, and the proposed \([[4,2,2]]\) RUS fault-tolerant construction.
  Error bars on the data are Wilson-score 95\% confidence intervals. At \(p_{\mathrm{phys}}=10^{-4}\), zero failures in \(50{,}000\) trajectories yield the displayed upper bound. The fitted slope of \(2.04\) for the fault-tolerant \texttt{CSWAP} is consistent with the second-order error scaling derived in Sec.~\ref{subsec:logicalerror}.
  }
  \label{fig:cswap_simulation}
\end{figure}

\begin{figure*}[t]
  \centering
  \includegraphics[width=\linewidth]{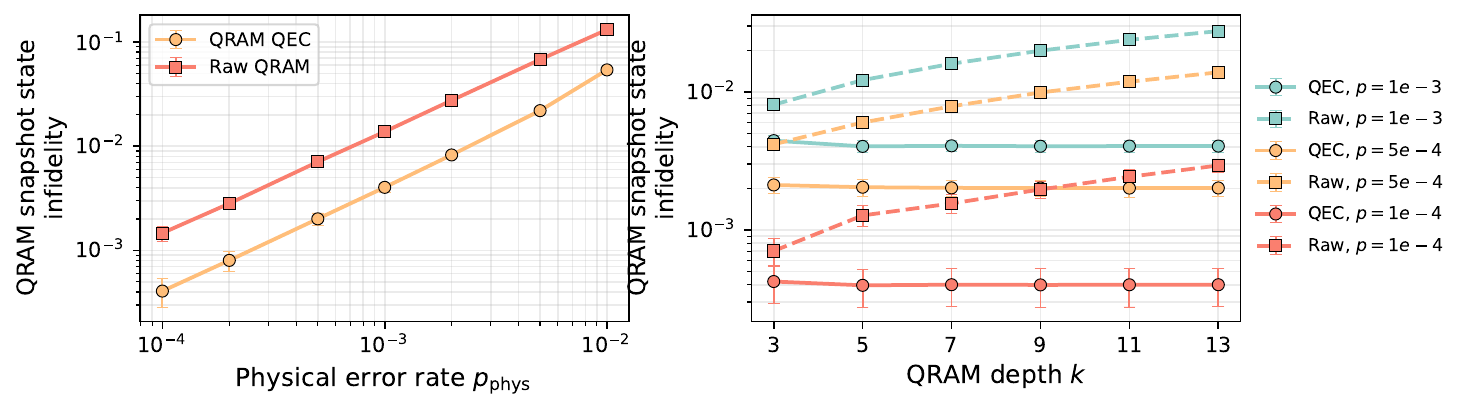}
  \caption{
  Snapshot infidelity \(1-F_{\mathrm{tree}}\) for valid BB routing states under the circuit-level bit-flip model.
  Left: raw and decoded infidelity versus \(p_{\mathrm{phys}}\) at fixed router-tree depth \(k=6\).
  Right: infidelity versus router-tree depth \(k=\log_2N-1\), with the root at level zero. The legend uses \(p=p_{\mathrm{phys}}\), with values \(10^{-3}\), \(5\times10^{-4}\), and \(10^{-4}\).
  Solid curves show the infidelity after QRAMQEC checks and MWPM decoding using the check outcomes and final-data parity detectors, while dashed curves show uncorrected BB routing.}
  \label{fig:qram_tree_state_simulation}
\end{figure*}

Figure~\ref{fig:cswap_simulation} compares the raw dual-rail routing operation, the non-fault-tolerant encoded construction of Fig.~\ref{fig:cswapov}, and the proposed \([[4,2,2]]\) RUS logical gate. The first two have the expected linear dependence on \(p_{\mathrm{phys}}\), with log-log slopes of \(1.00\). In contrast, the protected data give an effective slope of \(2.04\), consistent with the recovery of single reported erasures and a leading contribution from pairs of faults.

Figure~\ref{fig:qram_tree_state_simulation} shows the corresponding QRAMQEC results. Without decoding, the router-state infidelity increases with tree depth as more router locations contribute faults. With QRAMQEC, the depth dependence is substantially reduced over the simulated range because typical single faults in the interior are identified by the overlapping checks and corrected by MWPM. At \(k=6\) and \(p_{\mathrm{phys}}=10^{-3}\), decoding reduces \(1-F_{\mathrm{tree}}\) from approximately \(1.38\times10^{-2}\) to \(4.03\times10^{-3}\), a factor of \(3.4\). The remaining depth dependence is not uniform across the tree: the leaf routers retain a first-order contribution, while single root faults are corrected in the exact-syndrome model. The following estimate separates these boundary and interior contributions.

\subsection{Scaling of the router-state infidelity}
The circuit-level simulations above evaluate a noisy syndrome-extraction round followed by destructive readout. To understand the spatial scaling more directly, consider a simpler snapshot model in which bit flips accumulate on a valid BB routing state and are followed by an exact syndrome and minimum-weight correction. This deliberately isolates the protection supplied by the routing constraints themselves.

For a decoded snapshot, define
\begin{equation}
F_{\mathrm{tree}}
=
\sum_{\ell\in L_{\mathrm{good}}} p(\ell),
\end{equation}
where \(p(\ell)\) is the probability weight of leaf \(\ell\), taken to be
uniform in our simulations. Here each leaf is a level-\(k\) router, and
\(p(\ell)\) sums the address probabilities of its two memory cells.
\(L_{\mathrm{good}}\) is the set of leaf routers
whose complete root-to-leaf paths remain correct after decoding. Thus,
\(1-F_{\mathrm{tree}}\) is the total probability weight of paths containing
at least one residual routing error.

We consider an i.i.d.\ bit-flip channel on each physical rail,
\(\rho \mapsto (1-\epsilon)\rho+\epsilon X\rho X\).
Following the locality principle underlying the noise-resilience analysis of
BB QRAM in Ref.~\cite{hann}, a fault at a router can affect only the leaves
in its subtree. For uniformly weighted addresses, a router at
depth \(d\) therefore carries weight \(2^{-d}\), while there are \(2^d\)
routers at that depth. The total contribution of one tree level is thus
\(O(1)\), so a tree with \(k+1=\log_2 N\) router levels accumulates error with its
depth rather than with the total router count \(N-1\). Without decoding, the residual
error probability of each router is \(O(\epsilon)\), giving
\begin{equation}
\mathbb{E}[1-F_{\mathrm{tree}}]
=
O(k\epsilon).
\end{equation}

QRAMQEC suppresses the leading contribution of interior errors. Each
interior rail participates in two overlapping checks, so a single \(X\)
fault produces a defect pair and is corrected by MWPM at leading order.
Consequently,
\begin{equation}
\Pr[u\ \mathrm{bad}]
=
O(\epsilon^2)
\qquad
\text{for interior routers}.
\end{equation}
For \(k\ge2\), the single-fault syndromes of the root rails are unique,
so root faults are also corrected at leading order. The two rails of each
leaf router have the same single-fault syndrome. This ambiguity produces
the \(O(\epsilon)\) leaf-boundary contribution. Applying the same
subtree-weighting argument therefore gives
\begin{equation}
\mathbb{E}[1-F_{\mathrm{tree}}]
\le
c_1\epsilon+c_2 k\epsilon^2,
\label{eq:snapshot_scaling_style}
\end{equation}
For sufficiently small \(\epsilon\), the constants \(c_1\) and \(c_2\) can be chosen independently of \(k\): the decoding graph has bounded degree, and a residual path through a non-leaf rail requires at least two faults. Appendix~\ref{app:interior-bound} gives the path-counting bound. Thus QRAMQEC preserves
the intrinsic logarithmic size dependence of BB routing while suppressing
the depth-dependent contribution of interior errors from first to second order.

\section{Resource accounting}
\label{sec:resources}
We now quantify the overhead of the two protection mechanisms evaluated in the previous section: one protected logical \texttt{CSWAP} and one QRAMQEC syndrome-extraction round. We report qubit counts, operation counts, and circuit depth where they are fixed by the constructions, and separate these from recovery costs that depend on hardware-specific timing assumptions.

\subsection{Protected logical \texttt{CSWAP}}
The protected \texttt{CSWAP} gate acts on three \([[4,2,2]]\) blocks and therefore uses 12 data qubits. In the absence of faults, the logical operation consists of eight physical \texttt{CSWAP} steps. The additional cost comes from first-half reports that require undoing the executed gates and starting another attempt. We count executions of the original eight physical \texttt{CSWAP}s, including restarts, and list gates used for recovery separately.

Let
\begin{equation}
r=24\epsilon_0+36\epsilon_1+48\epsilon_2
\end{equation}
denote the first-order probability that one four-step first-half attempt is restarted. For the unbounded RUS protocol, the expected number of such attempts is \(1/(1-r)\). Conservatively counting all four steps of the second half gives the following upper estimate for the average number of physical \texttt{CSWAP} executions:
\begin{equation}
\overline S_{\texttt{CSWAP}}
\approx
4+
\frac{4-36\epsilon_0-54\epsilon_1-72\epsilon_2}
     {1-24\epsilon_0-36\epsilon_1-48\epsilon_2}.
\label{eq:cswap-resource-steps}
\end{equation}

As expected, \(\overline S_{\texttt{CSWAP}}=8\) in the zero-noise limit. This is an upper estimate because a second-half recovery can complete the operation before all four remaining gates are executed.

Each step of the original gate sequence is followed by erasure checks on all 12 data qubits. The fault-free path therefore produces 96 check outcomes, while the average number of check outcomes is at most \(12\overline S_{\texttt{CSWAP}}\). Although the 12 checks within one round can be performed in parallel, the next circuit step depends on their outcomes. Each physical \texttt{CSWAP} step therefore requires one wait for the check results and classical feedforward.

For the two-attempt version used to bound the circuit length, the probability that no third attempt is required is
\[
1-r^2+O(\epsilon^3).
\]
This quantity does not include other reported aborts or incorrect accepted outputs. With at most two attempts, the original gate sequence requires at most 12 physical \texttt{CSWAP} executions and 144 erasure-check outcomes.

A reported event can also invoke a branch-dependent correction. Depending on its location, this requires zero or one measurement of the control value, at most four additional gates to undo or complete the swap, and between one and three local \([[4,2,2]]\) recoveries. When the measured control value is known, recovery uses at most four physical \texttt{SWAP}s; an erasure on a nonparticipating physical qubit in \(C\) can instead require up to four additional physical \texttt{CSWAP}s. These correction operations are not included in Eq.~\eqref{eq:cswap-resource-steps}. Their exact average cost depends on the probability of a report at each location, so we keep them separate from the count of the original gate sequence.

To express the latency separately from hardware-specific gate times, let \(\tau_{\mathrm C}\), \(\tau_{\mathrm E}\), and \(\tau_{\mathrm F}\) denote the durations of a physical \texttt{CSWAP}, one parallel erasure-check round, and its classical feedforward, respectively. The fault-free critical path is
\begin{equation}
T_{\mathrm{FT},0}
=
8(\tau_{\mathrm C}+\tau_{\mathrm E}+\tau_{\mathrm F}),
\end{equation}
while the two-attempt limit allows at most 12 such steps.

A reported event adds a branch-dependent correction time \(T_{\mathrm{corr}}\), including control-value measurement when needed, reset, the gates used to undo or complete the swap, and local block recovery. The correction time \(T_{\mathrm{corr}}\) and the recovery-ancilla and reset counts depend on the hardware implementation.

\subsection{Error and latency tradeoff}
\label{subsec:resource-tradeoff}

To compare the benefit of correction with its additional waiting time, we
consider the same eight-gate encoded circuit with and without erasure checks.
For this estimate we set \(\tau_{\mathrm F}=0\); the latency expression above retains it explicitly.
Both circuits use six active before/after-gate locations and nine
idling locations per physical \texttt{CSWAP}, as in our analysis
of Fig.~\ref{fig:errormodel}. Set \(\epsilon_0=p\),
\(\epsilon_1=r_{\mathrm i} p\), and \(\epsilon_2=r_{\mathrm m} p\). A parallel check takes
\(w=\tau_{\mathrm E}/\tau_{\mathrm C}\) gate times and adds idle-erasure
probability at most \(r_{\mathrm i} wp\) on each of the 12 data qubits. The total
elementary fault probability per gate/check step is then bounded by \(B_{\mathrm{step}}p\),
where
\begin{equation}
B_{\mathrm{step}}=6+9r_{\mathrm i}+12r_{\mathrm i} w+12r_{\mathrm m}.
\label{eq:resource-step-exposure}
\end{equation}

We bound the protected failure probability by counting fault pairs.
Suppose the one-fault recovery condition holds, and every full conditional
correction has fault probability at most \(B_{\mathrm{rec}}p\), uniformly
over its branches. A two-attempt execution has at most 12 steps of the original gate sequence
and two correction segments. Padding shorter executions with unused
independent fault locations gives a fault-pair bound for these steps
\((12B_{\mathrm{step}}p)^2/2\). A faulty correction must follow a reported erasure in one of these steps, giving
at most \((12B_{\mathrm{step}}p)(2B_{\mathrm{rec}}p)\). Thus
\begin{equation}
P_{\mathrm{bad}}
\leq A p^2,\qquad
A=72B_{\mathrm{step}}^2+24B_{\mathrm{step}}B_{\mathrm{rec}},
\label{eq:resource-pair-budget}
\end{equation}
where \(P_{\mathrm{bad}}\) counts either an incorrect output or a reported
abort. This deliberately counts benign fault pairs as potentially harmful.

Without correction, an erased qubit remains outside the computational subspace, so the same
eight-gate circuit has error probability
\begin{equation}
P_{\mathrm{nonFT}}
=
1-(1-p)^{48}(1-r_{\mathrm i} p)^{72}.
\label{eq:resource-common-baseline}
\end{equation}
Comparing Eqs.~\eqref{eq:resource-pair-budget} and
\eqref{eq:resource-common-baseline} gives a sufficient physical-error regime
in which the protected circuit improves on the same encoded circuit without
checks. Because the protected bound counts benign fault pairs as potentially
harmful, the actual useful region can be larger.

For example, at \(p=10^{-4}\), \(r_{\mathrm i}=0.1\), \(r_{\mathrm m}=w=1\), and
\(B_{\mathrm{rec}}=60\), the protected failure bound is
\(5.80\times10^{-4}\), compared with \(5.51\times10^{-3}\) without checks.
The fault-free protected duration is twice the unchecked duration for
\(w=1\) and \(\tau_{\mathrm F}=0\), before additional delay from restarts and recovery is included.
Thus the error suppression comes with a corresponding latency cost.

\subsection{QRAMQEC syndrome extraction}

For \(N=2^n=2^{k+1}\) memory cells, the router tree occupies levels
\(0,\ldots,k\) and contains
\begin{equation}
R=N-1=2^{k+1}-1
\end{equation}
routers. Of these, \(N/2\) are leaf routers connected to the memory cells,
and \(N/2-1\) have router children. The dual-rail bit-flip construction
uses two data qubits per router, giving \(2N-2\) data qubits in total.

Each router with router children contributes two weight-three checks, so one extraction
round contains
\begin{equation}
M=2(N/2-1)=N-2=R-1
\label{eq:qec-check-count}
\end{equation}
check outcomes. A fully parallel implementation therefore uses \(M\) syndrome
ancillas, \(M\) ancilla resets, \(3M=3N-6\) data-to-ancilla
\texttt{CNOT}s, and \(M\) ancilla measurements.

The check--data incidence graph has degree three at each syndrome ancilla and
degree at most two at each data qubit. Its \texttt{CNOT}s can therefore be
scheduled in three interaction layers, followed by one parallel ancilla
measurement layer. Using fewer ancillas requires measuring the checks in more sequential groups.

\begin{table}[t]
\caption{\label{tab:qec-resources}
Resources for one fully parallel dual-rail QRAMQEC extraction round. The
\(k=6\) and \(k=13\) columns correspond to the representative tree depths
used in Fig.~\ref{fig:qram_tree_state_simulation}, with \(N=128\) and \(N=16{,}384\) memory cells, respectively.}
\begin{ruledtabular}
\begin{tabular}{lrrr}
Resource & \(N=2^{k+1}\) & \(k=6\) & \(k=13\)\\
\hline
Routers & \(N-1\) & 127 & 16,383\\
Dual-rail data qubits & \(2N-2\) & 254 & 32,766\\
Checks / syndrome ancillas & \(N-2\) & 126 & 16,382\\
Total data + syndrome ancillas & \(3N-4\) & 380 & 49,148\\
\texttt{CNOT}s per round & \(3N-6\) & 378 & 49,146\\
\texttt{CNOT} interaction layers & 3 & 3 & 3\\
\end{tabular}
\end{ruledtabular}
\end{table}

The counts in Table~\ref{tab:qec-resources} are outer-code counts. In a
concatenated implementation, the two rails of each router occupy the two
logical qubits of one \([[4,2,2]]\) block, giving
\[
4R=4N-4
\]
physical data qubits. The outer checks must then be implemented on encoded
rails with the corresponding inner-code protection, so the bare
\texttt{CNOT} and ancilla counts in the table are not the gate counts of a
fully protected extraction circuit.

The corresponding MWPM graph has \(O(N)\) vertices and edges. Its two
tree components can be decoded independently; the resulting classical
latency depends on the decoder implementation and is not included in the
hardware-independent resource counts above.

\section{Discussion}
\label{sec:discussion}
\subsection{Concatenating the two codes}
\label{sec:compose}

Our inner \([[4,2,2]]\) layer with fault-tolerant \texttt{CSWAP} protects individual routing operations against erasure faults, while the outer QRAM code uses correlations between neighboring routers to diagnose residual \(X\)-type errors across the tree. A natural question is whether these two layers of protection can be combined to provide a concatenated protection scheme for the entire QRAM query. A single \([[4,2,2]]\) block contains two logical qubits and can therefore encode both rails of one router, with
\[
W,L,R
\;\longleftrightarrow\;
\ket{00}_L,\ket{10}_L,\ket{01}_L.
\]
The outer QRAM checks then become logical parity measurements on the inner blocks. For example,
\begin{align}
S_{u,L}&=\bar Z_{u,1}\bar Z_{u_L,1}\bar Z_{u_L,2}
       =Z_{u,1}Z_{u,2}Z_{u_L,2}Z_{u_L,3},\nonumber\\
S_{u,R}&=\bar Z_{u,2}\bar Z_{u_R,1}\bar Z_{u_R,2}
       =Z_{u,1}Z_{u,3}Z_{u_R,2}Z_{u_R,3}.
\label{eq:concatenated_checks}
\end{align}
These weight-four parity checks follow from the logical operators of Sec.~\ref{sec:background}. Their measurement must also be protected against erasures by the inner code.

After address loading and before the bus is routed through the tree, we measure one round of the outer parity checks from Sec.~\ref{sec:qramcode} and decode the resulting syndrome. Since all valid routing patterns lie in the common \(+1\) eigenspace of these checks, the measurements preserve their superpositions while detecting errors that change the check parities. After decoding, however, the inferred \(X\) corrections cannot always remain in an ordinary Pauli frame. A later \texttt{CSWAP} is non-Clifford, so propagating an unresolved Pauli correction through it can produce a non-Pauli operation. During a QRAM query, the router correction must either be applied physically before the next non-Clifford gate or tracked by a compiler that handles non-Pauli corrections.

\subsection{Residual noise after an accepted \texttt{CSWAP}}
\label{subsec:compose_interface}

The state-selective biased erasure model leaves a residual \(X\) component even when no erasure is flagged. To describe this component and its effect on the concatenated scheme, consider the state-selective erasure channel
\begin{align}
& \mathcal{E}_{\mathrm{be}}(\rho)
  = K_0 \rho K_0^\dagger + K_1 \rho K_1^\dagger, \\
& K_0 = \ket{-}\!\bra{-}
  + \sqrt{1-p_e}\ket{+}\!\bra{+}, \qquad
  K_1 = \sqrt{p_e}\ket{e}\!\bra{+}.
\end{align}
Its no-flag Kraus operator can be written as
\begin{equation}
K_0=aI+bX,\qquad
a=\frac{1+\sqrt{1-p_e}}{2},\quad
b=\frac{\sqrt{1-p_e}-1}{2}.
\end{equation}

The no-flag operator \(K_0=aI+bX\) contains a coherent \(I\)-\(X\) superposition. Applying a physical \(Z\)-twirl at each elementary noise location removes these cross terms and converts the branch into a stochastic bit-flip channel. Conditioned on the twirled no-flag branch, the exact bit-flip probability is
\begin{equation}
\epsilon_X
=\frac{b^2}{a^2+b^2}
=\frac{p_e^2}{16}+O(p_e^3).
\label{eq:effective_inner_x}
\end{equation}

Randomized compiling can tailor coherent errors into stochastic Pauli noise~\cite{wallman2016noise} and has also been studied for QRAM~\cite{mehta2024analysis}. A \(Z\) on a physical \texttt{CSWAP} control commutes with the gate and may be tracked classically. A target \(Z\), however, propagates through \texttt{CSWAP} into a non-Pauli controlled operation. Target-qubit twirling therefore requires either physical gates or tracking of the resulting non-Pauli corrections.

We evaluate \(K_0\) at every noise location with ideal flag readout for all 64 logical computational-basis inputs of the three blocks. The logical circuit contains 48 active before/after locations and 72 idle locations. Physical \texttt{CSWAP} preserves total computational-basis parity, whereas a single physical \(X\) reverses it. All encoded three-block basis states have even parity, so inserting one \(X\) at any of the 120 locations gives zero projection onto the final code space. Under the physical twirl, the conditional entanglement infidelity equals the probability of being outside the code space to leading order in \(\epsilon_X\),
\begin{equation}
1-F_e=P_{\perp}+O(\epsilon_X^2)
=48\epsilon_{X,g}+72\epsilon_{X,i}+O(\epsilon_X^2),
\label{eq:no_flag_interface_audit}
\end{equation}
where \(\epsilon_{X,g}\) and \(\epsilon_{X,i}\) are the conditional bit-flip probabilities at active and idle locations. Here \(P_{\perp}\) is the probability of being outside the \([[4,2,2]]\) code space but within the physical qubit Hilbert space.

The inner stabilizer measurement detects the component outside the code space, and we use its syndrome for recovery whenever it determines a unique correction. If the syndrome is ambiguous, the protocol aborts. After recovery, the outer checks act on the logical rails of the \([[4,2,2]]\) blocks. For an accepted operation \(\mathcal A\), the acceptance probability is \(s(\rho)=\operatorname{Tr}\mathcal A(\rho)\) and the output is \(\mathcal A(\rho)/s(\rho)\). The elementary twirled no-flag map has \(s=a^2+b^2\), independent of the input, giving Eq.~\eqref{eq:effective_inner_x}. To evaluate the outer decoder, we use the stochastic logical-noise model
\begin{equation}
\begin{aligned}
\mathcal{R}_{\mathrm{acc}}(\rho)
={}&(1-q_1-q_{\mathrm{corr}}-q_{\mathrm{other}})\rho\\
&+\sum_i q_i X_i\rho X_i
+\sum_{\lvert S\rvert\geq 2}q_S X_S\rho X_S\\
&+q_{\mathrm{other}}\mathcal{N}_{\mathrm{other}}(\rho),
\end{aligned}
\label{eq:accepted_residual_channel}
\end{equation}
where
\[
q_1=\sum_i q_i,
\qquad
q_{\mathrm{corr}}=\sum_S q_S.
\]
Here the probabilities refer to logical errors after inner recovery, whereas \(\epsilon_X\) in Eq.~\eqref{eq:effective_inner_x} refers to elementary physical errors. The operators \(X_S\) represent correlated logical bit flips generated by one accepted logical gate, while \(\mathcal{N}_{\mathrm{other}}\) collects residual noise that is not described by the outer \(X\)-error model.

If inner-code syndrome measurement and recovery also satisfy the same one-fault recovery condition, all of these residual contributions begin at second order in the physical error rate. Their form depends on propagation through the routing gate, as illustrated by
\[
U_{\texttt{CSWAP}}X_CU_{\texttt{CSWAP}}^\dagger
=
X_C\mathrm{SWAP}_{AB},
\]
so propagation through the non-Clifford gate can produce non-Pauli errors.

\subsection{Error scaling of the concatenated scheme}
\label{subsec:combined_bound}

Let \(\epsilon_{\mathrm{snap}}\) denote the local bit-flip rate accumulated before the outer syndrome measurement. Including the independent logical-\(X\) residuals from accepted inner logical gates gives
\[
q=\epsilon_{\mathrm{snap}}+q_1.
\]
Here \(q_{\mathrm{corr}}\) is the probability of correlated \(X\) errors affecting a fixed number of logical rails, and \(q_{\mathrm{other}}\) is the probability of other local errors.

Applying the snapshot argument of Sec.~\ref{sec:results} gives
\begin{equation}
\mathbb{E}[1-F_{\mathrm{tree}}]
\;\lesssim\;
c_1 q+c_2 k q^2+c_3 k(q_{\mathrm{corr}}+q_{\mathrm{other}}),
\label{eq:composed_snapshot_bound}
\end{equation}
where \(c_1,c_2,c_3\) are independent of the memory size \(N=2^{k+1}\). The leaf boundaries contribute \(c_1q\), while independent faults in the interior contribute \(c_2kq^2\) after decoding. The final term accounts for correlated and non-\(X\) residual errors.

If protected storage and routing reduce the local residual error rate after inner recovery to
\[
q=O(p^2),
\]
then independent interior \(X\) faults contribute \(O(kp^4)\), while the leaf-boundary contribution is \(O(p^2)\). Writing \(\delta=q_{\mathrm{corr}}+q_{\mathrm{other}}\), the combined scaling is
\[
\mathbb{E}[1-F_{\mathrm{tree}}]
=O(p^2)+O(kp^4)+O(k\delta).
\]
Thus the depth-dependent contribution is fourth order when \(\delta=O(p^4)\). At finite bias, this regime is reached when the correlated and non-\(X\) residual rate is no larger than the independent two-error contribution.

Large-scale simulation of the full concatenated circuit is computationally demanding because the repeated non-Clifford \texttt{CSWAP} gates take the evolution beyond a stabilizer description. The separate trajectory and stabilizer simulations resolve the local gate dynamics and the spatial decoding, respectively. The concatenated construction combines these two forms of protection: the inner layer suppresses the local error rate through erasure recovery, while the outer layer uses the routing constraints to remove the leading contribution of independent interior bit flips.

\section{Conclusion}

We introduced a low-overhead protection framework tailored to bucket-brigade QRAM. A \([[4,2,2]]\) RUS logical \texttt{CSWAP} corrects single erasures during the execution of the logical gate, suppressing the leading error probability from first to second order. QRAMQEC uses the intrinsic redundancy of valid routing states to detect and correct interior bit flips through local parity checks and MWPM, without adding data qubits for redundancy. Quantum-trajectory and stabilizer simulations demonstrate quadratic error suppression for the protected gate and improved depth scaling of the router-state infidelity, respectively. Together, these results show how known erasure locations and intrinsic routing constraints enable low-overhead error correction in bucket-brigade QRAM.

\section*{Acknowledgments}
This material is based on work supported in part by the National Science Foundation under award CCF-2312754, in part by the Air Force Office of Scientific Research under award number FA9550-21-1-0209, and in part by the U.S. Department of Energy, Office of Science, National Quantum Information Science Research Center, Co-design Center for Quantum Advantage (C2QA) under Contract No. DE-SC0012704. YD acknowledges partial support by NSF CCF-2338063, by Quantum CT (under NSF Engines award ITE-2302908), by AFOSR MURI (FA9550-26-1-B036), by Boehringer Ingelheim, and NSF NQVL-ERASE (under award OSI-2435244). External interest disclosure: SMG is a consultant for, and YD is a consultant and equity holder in, D-Wave Quantum, Inc.  SMG and YD are co-founders and equity holders in Toric Labs, Inc.

\section*{AI disclosure}
The scientific ideas, methodology, and writing of this work were developed by the authors. Generative AI tools assisted with simulation-code design and manuscript preparation, including grammar checking. The authors retain full responsibility for the scientific content, claims, and conclusions.

\appendix
\section{Target-block erasure correction in the \texttt{CSWAP} protocol}
\label{app:target-erasure-case}

\begin{figure*}[t]
\centering
   \includegraphics[width=0.95\textwidth]{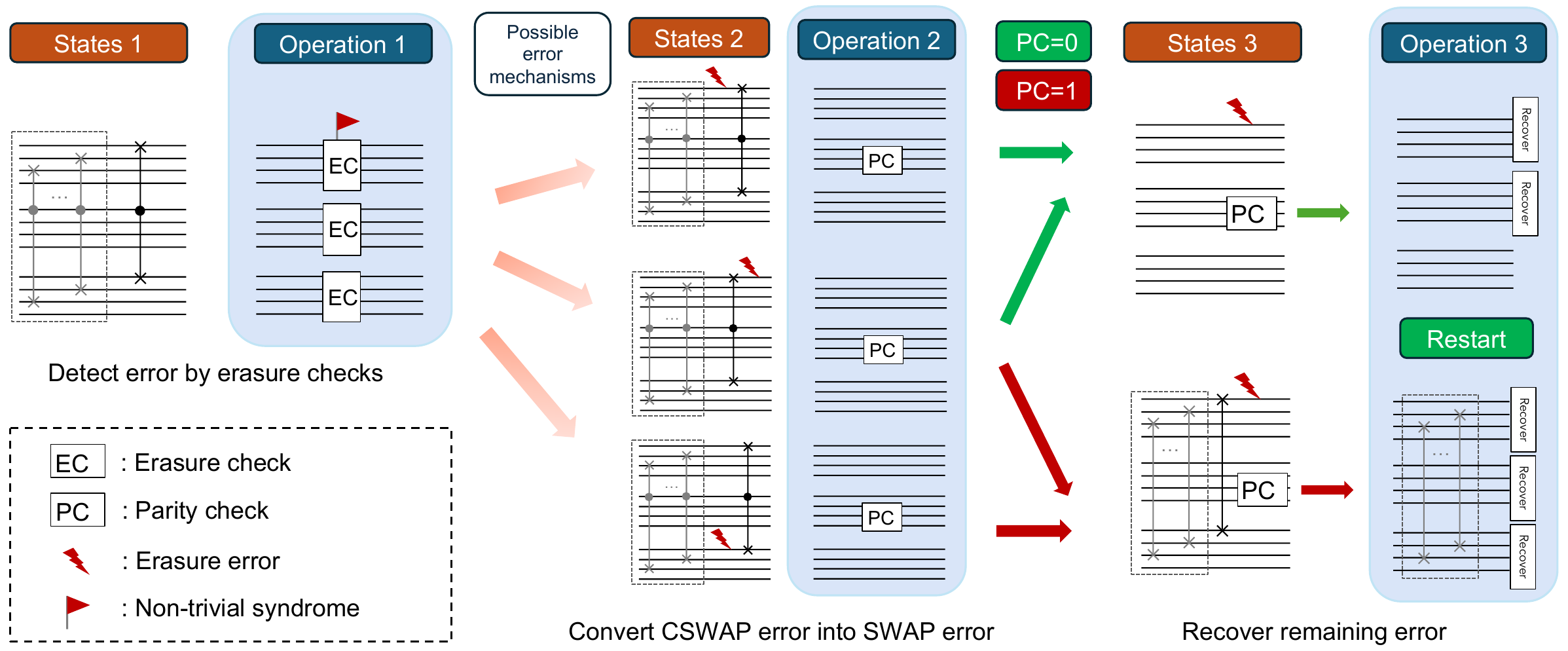}
\caption{Correction flow for a reported erasure in one of the target blocks during the first half of the logical \texttt{CSWAP} gate (Zone 2). Operations 1--3 detect the erasure, measure the active control qubit (PC=0 or 1), and undo the executed swaps as needed before local recovery and restart. States 1--3 show the state of the circuit before detection, after the erasure report, and after obtaining the measured control value, respectively.}
\label{fig:target-erasure-correction}
\end{figure*}

When the erasure checks report an erasure in one of the target blocks \(A\) or \(B\) during the first half (Zone 2) or the second half (Zone 4) of the logical gate, the affected data qubit may have been coherently transported by the partially executed sequence of physical \texttt{CSWAP} gates while the control block \(C\) remains in superposition. Once an erasure is detected, however, the check outcome identifies the erased target qubit and projects the state onto the subspace with an erasure at that location. As in the control-block case, the first step is to obtain the measured control value \(b\), which determines whether the preceding physical \texttt{CSWAP} gates have exchanged their targets.

Although no erasure is reported in \(C\), we still obtain \(b\) by measuring the active physical control qubit \(c_p\). The erasure report alone does not determine whether the preceding gates exchanged their targets, so this measurement tells us which swaps are needed to undo or complete the logical gate before local recovery. If \(b=0\), no additional swaps are required. If \(b=1\), we execute physical \texttt{SWAP}s on the relevant pairs \((A[j],B[j])\). For a first-half report (Zone 2), we undo the effect of all \texttt{CSWAP} gates executed up to and including the current step \(i\) by swapping the corresponding set of pairs. For a second-half report (Zone 4), we instead execute \texttt{SWAP} operations for the remaining \texttt{CSWAP} gates after step \(i\), thereby completing the intended output. Because measuring \(c_p\) dephases this physical qubit, we replace it and run \([[4,2,2]]\) erasure recovery on \(C\). Unlike the control-block case with an erasure reported on \(c_p\), no deliberate erasure of the current target pair is required here: the most recent physical \texttt{CSWAP} gate is unambiguously conditioned on the measured control value \(b\).

After these swaps and recovery of the measured control qubit, we apply \([[4,2,2]]\) erasure recovery to the affected target block. A single target-block report therefore requires two local block recoveries, one on \(C\) after measuring the control value \(b\), and one on the affected target block. In Zone 2, the protocol returns to step 0 and repeats the full RUS procedure. In Zone 4, it proceeds directly to the final state (step 8) without repetition.

\section{Logical \texttt{CSWAP} error-rate derivations}
\label{sec:appendix}
\label{app:cswap-derivations}

This appendix derives the logical error-rate estimates used in Sec.~\ref{subsec:logicalerror}, including faults in the erasure checks and recovery operations. We first classify pairs of faults according to whether they can be corrected, cause an abort, or remain undetected, and calculate their contributions for each half of the logical gate. We then account for early termination and RUS restarts when counting the expected number of executed gates, combine the two halves to obtain the leading logical error rate, and consider the limit of at most two attempts.

\subsection{Second-order error classification}
We use the error locations associated with each physical \texttt{CSWAP} step in Fig.~\ref{fig:errormodel} and include faults in the recovery operations triggered by an erasure report. To decide which fault pairs contribute to failure, we must consider what the reports tell us about the required recovery. If different fault histories give the same reports but require incompatible corrections, the protocol aborts.

Some pairs of faults can still be corrected if the reports provide enough information to choose the recovery. Two idling errors or two gate errors on different blocks can be corrected by applying \([[4,2,2]]\) recovery to each block. An erasure immediately after a gate can also be recovered if the following check gives a false negative but a later check detects it. An erasure missed by the final check, however, is an undetected failure. Two erasure-check errors on idle qubits or qubits participating in the gate can be corrected after deliberate erasure.

Detected failures produce erasure reports for which the recovery procedure cannot determine a unique correction. These include two erasures in the same block or a before-gate error plus an idling error on different blocks (possibly mixed with erasure-check errors). The probability of these detected failures during one physical \texttt{CSWAP} step is
\begin{align*}
    p_{det} =&  9 \epsilon_1^2 (\textrm{two idling})
    + 36 \epsilon_0\epsilon_1 (\textrm{one gate + one idling}) \\
    &+ 45 \epsilon_2^2(\textrm{two check errors})\\
    &+36\epsilon_1\epsilon_2(\textrm{check error + idling})\\
    &+18\epsilon_0\epsilon_2(\textrm{check error + gate}).
\end{align*}

Undetected failures include errors in the recovery circuit after a detected
error, and a gate error immediately following a false-negative erasure-check
outcome. Let \(m\) be the number
of physical \texttt{SWAP}s executed during recovery to undo or complete
the swap. The probability at one step is
\begin{align*}
    p_{\mathrm{undet}}(m)
    &= 3 \epsilon_0\epsilon_1
    + 3\epsilon_0(2\epsilon_2+2m\epsilon_0)
    + 9\epsilon_1 \cdot 3m\epsilon_0\\
    &=3 \epsilon_0\epsilon_1 + 6\epsilon_0\epsilon_2
    +6m\epsilon_0^2+27m\epsilon_0\epsilon_1,
\end{align*}
where \(2\epsilon_2\) comes from parity-check errors in recovery and
\(m\epsilon_0\) comes from faults in these recovery \texttt{SWAP}s. For step
\(i\in\{1,2,3,4\}\) of the repeatable first half, undoing the executed swaps includes the
current pair, so \(m_i^{(\mathrm{R})}=i\). For step \(i\) of the
second half, completing the swap requires only the pairs after the current
step, so \(m_i^{(\mathrm{O})}=4-i\). Therefore
\begin{equation}
\sum_i m_i^{(\mathrm{R})}=10,
\qquad
\sum_i m_i^{(\mathrm{O})}=6.
\end{equation}

Summing \(p_{\mathrm{det}}+p_{\mathrm{undet}}\) over each four-step half gives
\begin{align}
    p_{\mathrm{R}}
    &= 60\epsilon_0^2 + 36\epsilon_1^2 +180\epsilon_2^2 \nonumber\\
    &\quad+426\epsilon_0\epsilon_1
    + 96\epsilon_0\epsilon_2
    + 144\epsilon_1\epsilon_2, \label{eq:repeat_half_second_order}\\
    p_{\mathrm{O}}
    &= 36\epsilon_0^2 + 36\epsilon_1^2 +180\epsilon_2^2 \nonumber\\
    &\quad+318\epsilon_0\epsilon_1
    +96\epsilon_0\epsilon_2
    +144\epsilon_1\epsilon_2. \label{eq:once_half_second_order}
\end{align}

\begin{figure}[t]
    \centering
    \includegraphics[width=\linewidth]{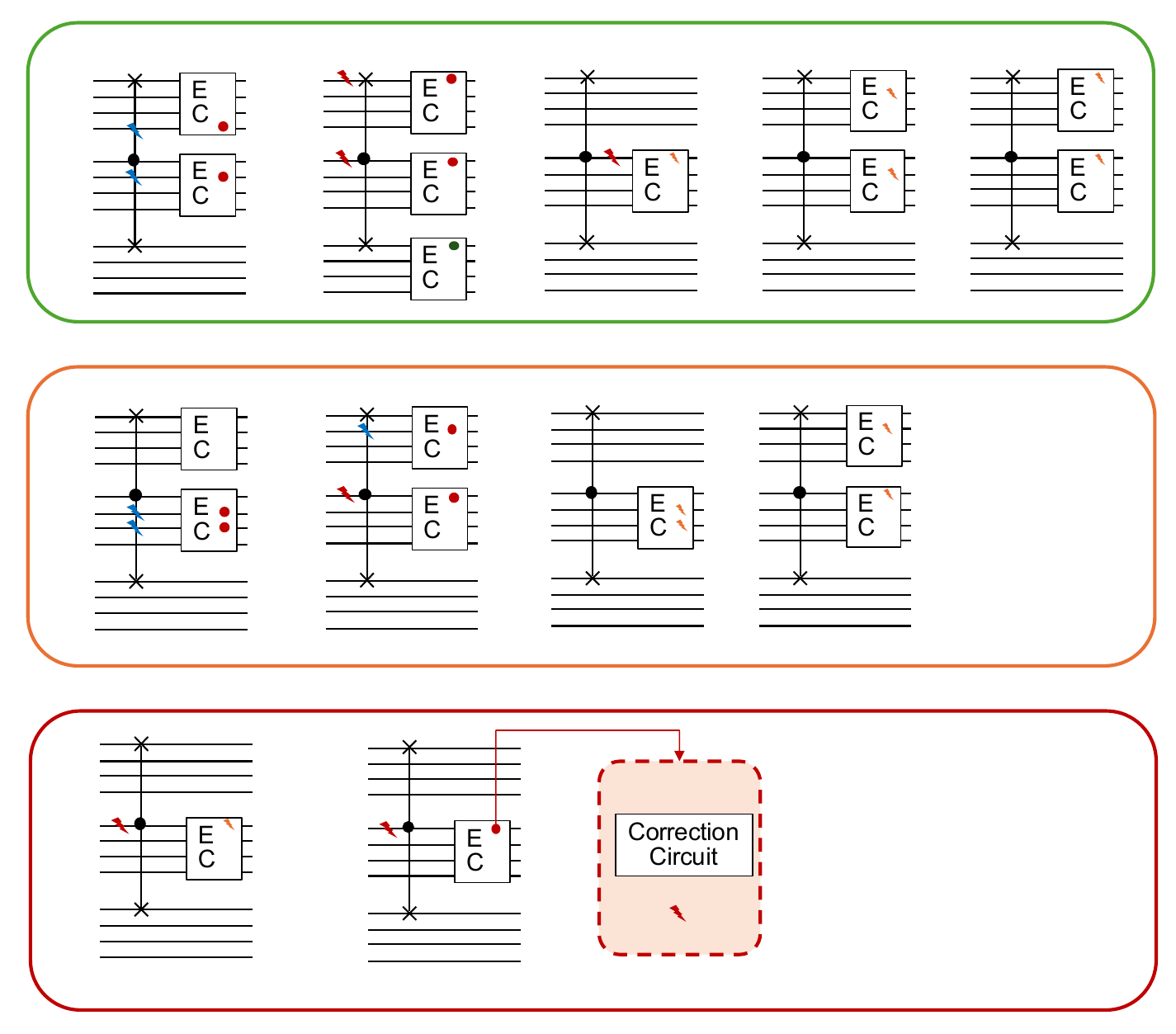}
    \caption{Examples of correctable errors (top), detected failures (middle), and undetected failures (bottom).}
    \label{fig:errexample}
\end{figure}

\subsection{Number of attempts}
For each physical \texttt{CSWAP} step $i$, let $f_i$ be the probability of an erasure report. The probability of no report is
\begin{align*}
g_i = 1 - f_i
= 1- 6\epsilon_0 - 9\epsilon_1 -12 \epsilon_2
+O(\epsilon^2).
\end{align*}
Each attempt aims to complete all four first-half steps without a report. This occurs with probability
\begin{align*}
P_{\textrm{success}} = \prod_{i=1}^{4} g_i.
\end{align*}
To first order, the expected number of attempts is
\begin{align*}
\mathbb{E}[\text{attempts}]
&= \frac{1}{P_{\text{success}}} 
 = \frac{1}{g_1 g_2 g_3 g_4} \\
&\approx \frac{1}{1 - 24\epsilon_0 - 36\epsilon_1 - 48\epsilon_2}.
\end{align*}

An attempt ends as soon as an erasure is reported, so it need not execute all four steps. Accounting for this early termination, the expected number of physical \texttt{CSWAP} steps executed in the repeatable first half is
\begin{align*}
\mathbb{E}[\text{steps}_{\mathrm{first}}]
&= \frac{1}{g_1g_2g_3g_4}
 + \frac{1}{g_2g_3g_4}
 + \frac{1}{g_3g_4}
 + \frac{1}{g_4} \\
&\approx
\frac{4 - 36\epsilon_0 - 54\epsilon_1 - 72\epsilon_2}
     {1 - 24\epsilon_0 - 36\epsilon_1 - 48\epsilon_2}
\quad (\text{to first order}).
\end{align*}
Each of the four steps in the second half is executed at most once after the first half succeeds. Counting all four gives the upper estimate
\begin{align*}
\mathbb{E}[\text{steps}_{\mathrm{total}}]
\le 4+\mathbb{E}[\text{steps}_{\mathrm{first}}],
\end{align*}
which gives eight physical \texttt{CSWAP} steps in the zero-noise limit.

\subsection{Overall logical error rate}
The overall logical error rate combines contributions from the repeatable first half and the second half:
\begin{align}
p & =  p_{\textrm{repeat}}+p_{\textrm{once}} \\
&= \sum_{i=1}^4
\mathbb{E}[\text{executions of repeatable step }i]\,
p_i^{(\mathrm{R})}
+p_{\mathrm{O}}.
\end{align}
Each \(p_i^{(\mathrm{R})}\) is already second order, and restarts increase the expected execution count above one by a first-order amount. Their product therefore changes the logical error rate only at third order. Using
Eqs.~\eqref{eq:repeat_half_second_order} and
\eqref{eq:once_half_second_order} gives
\begin{align}
p
&=p_{\mathrm{R}}+p_{\mathrm{O}}+O(\epsilon^3)\nonumber\\
&\approx 96\epsilon_0^2 + 72\epsilon_1^2 +360\epsilon_2^2 \nonumber\\
&\quad+744\epsilon_0\epsilon_1
+192\epsilon_0\epsilon_2
+288\epsilon_1\epsilon_2.
\label{eq:corrected_total_second_order}
\end{align}

\subsection{Effect of limiting RUS to two attempts}
For at most two RUS attempts, the probability that no third attempt is
required, considered separately from other failure outcomes, is
\begin{align*}
P_{\textrm{no third attempt}} = 1- (24\epsilon_0 + 36\epsilon_1 +48 \epsilon_2)^2,
\end{align*}
while the fault-pair estimate above remains
\begin{align}
p \approx {}&
96\epsilon_0^2 + 72\epsilon_1^2 +360\epsilon_2^2 \nonumber\\
&+744\epsilon_0\epsilon_1
+192\epsilon_0\epsilon_2
+288\epsilon_1\epsilon_2.
\end{align}

\section{Syndrome-extraction circuits for QRAMQEC}
\label{app:syndrome-circuits}

The QRAMQEC checks used for the bit-flip-biased model are \(Z\)-type parity checks on neighboring router rails.
For a check \(S=Z_{q_1}Z_{q_2}\cdots Z_{q_w}\), we use one syndrome ancilla initialized in \(\ket{0}\), apply CNOT gates sequentially from each data qubit \(q_i\) to the syndrome ancilla, and then measure the ancilla in the computational basis.
The measurement outcome gives the parity of the listed data qubits and hence the syndrome bit for \(S\).
This is the standard circuit for a \(Z\)-parity measurement, specialized to the local QRAMQEC checks of Sec.~\ref{subsec:bitflip}.

A circuit-level implementation can use a sequential CNOT schedule for each check. When two checks overlap on a data qubit, their CNOT gates are scheduled so that only one syndrome ancilla interacts with that qubit at a time. Checks on disjoint BB elements can be measured in parallel, while overlapping checks are serialized by layer or by an edge coloring of the check graph.

The CNOT orientation is useful for the biased error model because the dominant faults are \(X\)-type faults. With the data qubits as controls and the syndrome ancilla as the target, an \(X\) fault on the ancilla does not propagate back to the data. An \(X\) fault on a data qubit can propagate to the ancilla and change the reported syndrome, but it cannot spread through the ancilla to another data qubit. Syndrome-extraction faults therefore remain local in the QRAMQEC decoding graph under the biased model considered here.

The snapshot simulations in Sec.~\ref{sec:results} use one noisy ancilla-based extraction round. They include independent Pauli-\(X\) faults on data and ancilla idling locations, on both qubits after each extraction \texttt{CNOT}, and immediately before check and final-data measurements. The detector model also compares the final data parities with the last ancilla outcomes.

\section{Boundary contribution and cost tradeoff}
\label{app:boundary-checks}

The first-order boundary contribution in the exact-syndrome analysis comes from the leaf routers because their two rails have identical single-fault syndromes. For \(k\ge2\), each root rail has a unique single-fault syndrome and is corrected by MWPM. Virtual boundary vertices represent the single-defect events at both ends of the decoding graph.

A physical extension that duplicates leaf boundary rails could suppress part of the \(O(\epsilon)\) contribution, but there are \(O(N)\) leaves in a size-\(N\) memory. The added qubits and syndrome ancillas would therefore also cost \(O(N)\). We retain the open boundary to avoid this additional hardware cost.

The decoder weights can be adjusted to measured error rates without adding hardware. In our simulations, each edge ending at the virtual leaf boundary has the weight assigned to a single fault under the i.i.d.\ model. If measured error rates differ between boundary and interior operations, we can use those rates to set the matching weights. This adjustment accounts for the noise variation but leaves the code distance and \(O(\epsilon)\) boundary scaling unchanged.

\subsection{Interior error bound}
\label{app:interior-bound}

For the exact-syndrome, i.i.d.\ model with \(k\ge2\), represent each boundary edge as ending at a separate virtual leaf. The decoding graph is then a forest with check-vertex degree at most three. Let \(E\) be the actual error and \(C\) a minimum-weight correction. Their symmetric difference decomposes into boundary-to-boundary paths. At least half the edges on each path must belong to the actual error \(E\). If fewer than half did, toggling the correction along that path would reduce its weight, contradicting the choice of \(C\) as a minimum-weight correction.

A residual path containing a fixed non-leaf rail \(q\) has length \(L\ge3\). At most \(L2^{L-1}\) paths of length \(L\) contain that rail, and each requires at least \(\lceil L/2\rceil\) independent faults. A union bound therefore gives
\begin{equation}
\Pr[q\in E\mathbin{\triangle}C]
\le \sum_{L\ge3} L2^{2L-1}\epsilon^{\lceil L/2\rceil}
=O(\epsilon^2).
\end{equation}
For any fixed \(\epsilon_*<1/16\), this bound is uniform in \(k\) for \(0\le\epsilon\le\epsilon_*\). The same count for leaf rails starts at \(L=2\), giving \(O(\epsilon)\). Since the subtree weights sum to one at each level, summing over the \(k\) non-leaf levels and the leaf level yields Eq.~\eqref{eq:snapshot_scaling_style}.

\bibliography{apssamp}% Produces the bibliography via BibTeX.
\end{document}